\documentclass{JHEP3}

\title{Localized modes at a D-brane--O-plane intersection and heterotic Alice strings}

\author{Jeffrey A. Harvey and Andrew B. Royston \\ Enrico Fermi Institute and Department of Physics \\ 5640 Ellis Av.,
Chicago Illinois 60637, USA \\ E-mail: \email{harvey@theory.uchicago.edu}, \email{aroyston@uchicago.edu}}

\abstract{We study a system of $N_c$ $D3$-branes intersecting $D7$-branes and $O7$-planes in $1+1$-dimensions. We use anomaly cancellation and string dualities to argue that there must be chiral fermion zero-modes on the $D3$-branes which are localized near the $O7$-planes. Away from the orientifold limit we verify this by using index theory as well as explicit construction of the zero-modes. This system is related to F-theory on K3 and heterotic matrix string theory, and  the heterotic strings are related to Alice string defects in $\mathcal{N}=4$ Super-Yang-Mills. In the limit of large $N_c$ we find an $AdS_3$ dual of the heterotic matrix string CFT.}

\keywords{Anomalies in Field and String Theories, D-branes, AdS-CFT Correspondence, F-theory}

\preprint{EFI-07-24}

\usepackage{bbm}
\usepackage{amsmath}
\usepackage{slashed}

\begin{document}

\section{Introduction and motivation} \label{Introduction}

The AdS/CFT correspondence \cite{Maldacena,Gubser,Wittenads} can be generalized
to a duality between conformal field theories with defects and $D$-brane configurations
in Anti de Sitter space (AdS) which typically wrap AdS subspaces of the ambient AdS \cite{Karcha,Karchb,DeWolfe}. One important example of this duality arises by considering $Dp$-branes
which intersect $N_c$ $D3$-branes in the large $N_c$ limit. At weak 't Hooft coupling this
system is described by a defect in the $\mathcal{N}=4$ SYM conformal theory on the $D3$-branes. At
strong couping it is described by a geometry in which the $Dp$-brane wraps a subspace of the
$AdS_5 \times S^5$ near-horizon geometry of the $D3$-branes.

Such systems and their generalizations
to nonconformal theories play an important role in recent attempts to provide a string theoretic construction of
strong coupling dual descriptions of QCD, but in such models the full structure of the
correspondence has not yet been worked out. One of the motivations for the current
research was to work out the mapping between states and operators in a defect AdS/CFT system
with chiral fermions. Intersections with chiral fermions  occur in \cite{Sakai} which involves
$D8/D4$-brane intersections.  The model considered here, involving $D7/D3$ intersections,
is easier to analyze because of the more direct connection to AdS/CFT.  This example was mentioned in \cite{Skenderis}, but to our knowledge has not been studied in depth.  In fact the details
of this correspondence will appear elsewhere \cite{inprep} while here we focus on
some foundational material needed to understand the structure of this system from the
field theory point of view. The $D7/D3$ system has a rich web of connections to many
other well-studied systems in string theory including stringy cosmic strings, string defects
in $\mathcal{N}=4$ SYM, F-theory on K3 and its various dual descriptions, and the matrix theory
description of heterotic string theory. It  may also be useful in constructing the elusive
instanton corrections to the $O7$-plane anomalous couplings \cite{Julia}.

This paper is organized as follows. We start out in the following section by describing the system we will be studying and reviewing some of the relevant material on $D7$-branes and $O7$-planes and
the corresponding supergravity solutions. We then identify the obvious zero-modes on the
$1+1$-dimensional intersection and study the question of anomaly cancellation by anomaly
inflow. We are led to conclude from this analysis that there must be additional zero-modes
on the $D3$-branes which are localized near the intersection of the $D3$-branes and
$O7$-planes. We then establish the existence of these zero-modes first by constructing
the effective action on the $D3$-branes and using index theory, and then by explicit construction.  In Section 4 we use string dualities and the explicit formulae for the zero-modes to deduce the dependence of the Type I and heterotic string couplings on the 7-brane moduli.

Much of our analysis becomes trivial in a particular limit where the space transverse to the
$D7/O7$-planes is compact and the $D7$-branes and $O7$-planes coincide so that
the system can be studied using the standard perturbative string analysis of orientifolds. In
fact in that case the system is $T$-dual to a $D1$-string in Type I theory. Our analysis is however
more general and allows us to also analyze the system for a noncompact transverse space
and also away from the orientifold limit where one encounters regions of strong coupling.

\vspace{.3cm}

\section{Description of the system} \label{SystemDescription}

\subsection{Stacks of branes and $o$-planes}

We consider a transverse intersection of $D3$-branes with $D7$-branes and $O7$-planes.  The coordinate axes are taken such that the branes span the directions marked in Table \ref{table0}.
\TABLE{ \label{table0}
\begin{tabular}{l|c|c|c|c|c|c|c|c|c|c}
& 0 & 1 & 2 & 3 & 4 & 5 & 6 & 7 & 8 & 9 \\
\hline
$D7$, $O7$ & x & x & & & x & x & x & x & x & x \\
\hline
$D3$ & x & x & x & x & & & & & &  \\
\end{tabular}
\caption{Brane orientations}}

Bulk coordinates are denoted by $x^M$, $M = 0,\ldots, 9$ and are divided into $x^{M} = (x^m, y^\alpha)$ with $m = 0,\ldots, 3$ and $\alpha = 1,\ldots, 6$.  We also use indices $A,B$, $a,b$, and $\underline{\alpha},\underline{\beta}$ to denote corresponding tangent space directions.  We further divide the spacetime directions along the $D3$-brane into $x^m = (x^\mu, z, \bar{z})$ with $\mu = 0,1$, and $z = x^2 + i x^3$.  Corresponding tangent space indices are underlined.  Lightcone coordinates $x^{\pm} = x^0 \pm x^1$ will also be used.  We denote the
arbitrary number of $D3$-branes by $N_c$.

In the strict $g_s = 0$ limit, the number of $D7$-branes and $O7$-planes can be arbitrary, but as soon as $g_s \neq 0$ many of these configurations become inconsistent.  As we will review in the next section, one-half BPS 7-brane solutions in supergravity with arbitrary $g_s$ only exist for certain special numbers and combinations of $(p,q)$ type 7-branes.  Furthermore, as soon as $g_s \neq 0$, the full back reaction of the 7-branes on the metric, dilaton, and R-R scalar (axion) must be considered; there is no $\alpha' \rightarrow 0$ decoupling limit.  This is easily seen from the fact that Newton's constant in front of the IIB supergravity action and the $D7$-brane tension in front of the 7-brane DBI plus WZ action have the same powers of $\alpha'$.  Thus the solutions will not depend on $\alpha'$.

There have been many interesting studies of adding flavors to the classic AdS/CFT correspondence using 7-branes.  One can work in the strict $g_s = 0$ probe limit \cite{KarchKatz}, but if one is interested in subleading effects, the fully back reacted supergravity solution must be considered \cite{AFM,GranaPolchinski,BLZY}.  In general, the exact solution is not known.  Note, though, in the simplest 7-brane background--a $\mathbbm{Z}_2$ orientifold--a very explicit study of the correspondence including subleading effects can be made \cite{APTY}.  Even in the more general setting, one can make much progress for the following reason.  The $D3$-branes in all of these setups are parallel to the $D7$-branes and close to or coincident with them.  Therefore, one can go a long way by approximating the supergravity solution in a region near the $D7$-branes.

Our setup is very different, as the $D3$-branes are extended in the directions transverse to the 7-branes.  We will want to consider $g_s \neq 0$ as this leads to some interesting results, and therefore it will be crucial to work with the full 7-brane supergravity solution.  Hence, there will be restrictions on the number and type of 7-brane configurations we can consider.

In this paper we will work in a regime where the $D3$-branes are well described by an effective field theory on their worldvolume; in other words we assume $g_s N_c << 1$.

\subsection{Review of 1/2-BPS seven-brane solutions}

A static 7-brane sources the metric and axidilaton $\tau = C_0 + i e^{-\Phi} \equiv \tau_1 + i \tau_2$.  The supergravity equations of motion of this system, or alternatively the preservation of supersymmetry, require $\tau$ to be a holomorphic\footnote{ \label{footnote1}Whether $\tau$ must be a holomorphic or anti-holomorphic function depends on one's definition of positive orientation, or equivalently on whether one wishes the supersymmetries preserved by the 7-brane to have positive chirality or negative chirality.  Later, we will choose the supersymmetries preserved by the $D3$-branes to be in the $(\mathbf{2}, \bar{\mathbf{4}})$ of the corresponding  $SO(1,3) \times SO(6)$, and at the same time declare that the supersymmetries preserved by the $D3/D7$ intersection are right-handed with respect to the $SO(1,1)$ of the intersection.  This puts the four preserved (complex) supercharges in the $(-1/2,-1/2,\bar{\mathbf{4}})$ of $SO(1,1) \times SO(2) \times SO(6)$, so that the corresponding spinor parameterizing IIB supersymmetry variations, $\epsilon$, is in the $(1/2,1/2,\mathbf{4})$.  Requiring that the supersymmetry variation of the dilatino vanishes then implies that $\tau$ must be anti-holomorhpic.} function of $z$, and near the source it should behave as $\tau \sim \ln{(z - z_0)}$.  The key ingredient to having solutions with a dilaton that does not diverge as $z \rightarrow \infty$ and a finite energy per unit 7-brane volume, is to use the $SL(2,\mathbbm{Z})$ invariance of IIB string theory to allow the axidilaton to make jumps by $PSL(2,\mathbbm{Z})$ transformations.  Values of the dilaton that are related by such transformations are physically equivalent.  One can alternatively represent such solutions with a $\tau(z)$ that does not have discontinuous jumps by making use of Klein's modular j-function--a 1:1 and onto map $j: F_0 \rightarrow \hat{\mathbbm{C}}$, where $F_0$ is the fundamental domain of $PSL(2,\mathbbm{Z})$.
This map is holomorphic everywhere except at the cusps of $F_0$, at which $j(i) = 1$, $j(\rho) = 0$, and $j(i\infty) = \infty$, where $\rho = e^{2 i \pi/3}$.  Around these points, $j^{-1} : \hat{\mathbbm{C}} \rightarrow F_0$ behaves as
\begin{eqnarray} j^{-1}(z) &=& \left\{ \begin{array}{l} i + \sqrt{z - 1} + O((z-1)^1) \\
\rho + z^{1/3} + O(z^{2/3}) \\
\frac{i}{2\pi} \log{z} + O(1/z) \end{array} \right. \end{eqnarray}
respectively.  Around these points $\tau = j^{-1}(z)$ has $PSL(2,\mathbbm{Z})$ monodromy $S$, $T^{-1} S$, and $T$ respectively, where $S,T$ are the usual generators of $SL(2,\mathbbm{Z})$ and $\Lambda = $ {\scriptsize$\begin{pmatrix} a&b \cr c&d \end{pmatrix}$} $\in PSL(2,\mathbbm{Z})$ acts on $\tau$ via
\begin{eqnarray} \Lambda \tau & \equiv & \frac{a \tau + b}{ c \tau + d} \ . \label{tautrans} \end{eqnarray}

If we were simply to set $\tau(z) = j^{-1}(z)$, the $\log{z}$ behavior looks promising, but the failure of holomorphicity around $z = 0,1$ indicates that SUSY is not globally preserved.  This can be remedied by taking
\begin{eqnarray} \tau(z) & = & j^{-1}(f(z)), \end{eqnarray}
where $f(z) = P(z)/Q(z)$ is a meromorphic function with the properties
\begin{eqnarray} \begin{array}{l l} f(z) = 1 + (z - z_i)^2 + O((z-z_i)^4), & \textrm{near $z_i$ such that $f(z_i) = 1$} \ , \\
f(z) = (z - z_\rho)^3 + O((z - z_\rho)^6), & \textrm{near $z_\rho$ such that $f(z_\rho) = 0$ \ .} \end{array}  \label{frequirements} \end{eqnarray}
We also label the poles of $f$ with the notation $z_{i\infty}^{(n)}$; these points will correspond to locations of 7-branes.  We can assume that $P,Q$ are polynomials of the same degree so that $z = \infty$ is not a special point of $f$--this is just a choice of coordinate system.  If $f$ has $N_f$ poles, then the modular invariant and nowhere vanishing metric is given by
\begin{eqnarray} ds^2 & = & \eta_{\mu\nu} dx^\mu dx^\nu + e^{a(z,\bar{z})} dz d\bar{z} + \delta_{\alpha\beta} dy^\alpha dy^\beta \ , \end{eqnarray}
with
\begin{eqnarray} e^{a(z,\bar{z})} & = & \tau_2 \eta^{2}(\tau) \bar{\eta}^2(\bar{\tau}) \displaystyle\biggl[ \frac{1}{\prod_{n = 1}^{N_f} (z - z_{i \infty}^{(n)}) (\bar{z} - \bar{z}_{i \infty}^{(n)})} \displaystyle\biggr]^{1/12} \ . \end{eqnarray}
This is smooth everywhere except at the $z_{i\infty}^{(n)}$ where it behaves like $\log{|z-z_{i\infty}^{(n)}|}$, due to the factor of $\tau_2$.  (The explicit factors of $(z-z_{i\infty}^{(n)})^{-1/12}$ cancel out the zeroes of the Dedekind eta function at these points).  These solutions were first constructed in \cite{GSVY} as ``stringy cosmic strings,'' before the invention of $D$-branes, and they were later interpreted as $D7$-brane solutions by \cite{GGP}.  A recent analysis is given in \cite{Berg}, whose notation we follow.

The minimum number of poles of such an $f$ satisfying \eqref{frequirements} is $N_f = 6$.  $\tau(z)$ is then a 6:1 wrapping of $F_0$ by $\hat{\mathbbm{C}}$, holomorphic everywhere except at the $z_{i\infty}^{(n)}$.  Note that $6 = 3\cdot 2$ is the product of orders of monodromies around the cusp points $i,\rho$ of $F_0$.  Any $f(z)$ satisfying the above requirements must in fact have $N_f$ divisible by $6$.  On the other hand, since $\tau$ goes to a non-zero constant as $z \rightarrow \infty$, the metric behaves as
\begin{eqnarray} \lim_{|z| \rightarrow \infty} e^{a} & \sim & \frac{1}{|z|^{N_f/6}} \ . \end{eqnarray}
It follows that the transverse space is asymptotically conical, with deficit angle $\delta = (2\pi) \frac{N_f}{12}$.  Hence for $N_f = 12$ the space is asymptotically cylindrical and for $N_f > 12$ it becomes compact.  For the compact case, it turns out that only $N_f = 24$ avoids problems at $z = \infty$.  The transverse space has deficit angle $4\pi$--ie. it has become an $S^2$ (or $\mathbbm{CP}^1$).  Thus there are only three values\footnote{If one allows $\tau$ to have nontrivial monodromy around $z=\infty$, then it is possible to have other values of $N_f$ that are divisible by only 3 or 2.  However, this is accomplished by setting $z_\rho = \infty$ or $z_i = \infty$.  As a result, $\tau$ is everywhere constant and equal to $\rho$ or $i$.  Thus these solutions are nonperturbative in nature and we do not consider them.} $N_f$ can take, $6,12,24$.

Around the $z_{i \infty}^{(n)}$, $\tau$ has monodromy $T$, so that $C_0 \rightarrow C_0 + 1$.  Naively then, these points correspond to the locations of ordinary $D7$-branes, on which $F1$-strings (or $(p,q) = (1,0)$ strings) end, as they have magnetic R-R 0-form charge 1.  However, in constructing these solutions we have essentially gauged $SL(2,\mathbbm{Z})$, which acts on the $(p,q)$ charges of $F1/D1$ bound states, so this naive intuition turns out to be wrong.  Before discussing this, we mention that \cite{Berg} has argued that the orbifold points of $\tau(z) = j^{-1}(z)$ at $z=0,1$  correspond to the location of (p,q) 7-branes in a different conjugacy class--i.e. not related to the ordinary (1,0) 7-branes by an $SL(2,\mathbbm{Z})$ transformation.  If so, it is possible to construct solutions with any value of $N_f$.  However, some of these branes, if they exist at all, have negative mass, and their worldvolume dynamics is not well understood, so we will not consider them.

In fact, it is clear that one can not have 24 $D7$-branes on a compact (transverse) space, as there would be nowhere for the field lines to go.  If some of the 7-branes are more general (p,q) 7-branes then we might avoid this inconsistency.  By observing that $\tau(z)$ is an elliptic fibration over a $\mathbbm{CP}^1$ base, (ie. compactification of F-theory on K3), and noting that such a fibration has only 18 relevant complex moduli to vary, Vafa \cite{Vafa} argued that there are only 18 relative positions\footnote{For the configurations we will consider, the $18$ complex moduli are accounted for as follows.  We have the positions of 4 $O7$-planes and 16 $D7$-branes, as well as the asymptotic value of the axidilaton, giving $21$ free complex parameters.  However, 3 of the $7$-brane positions can be fixed arbitrarily using the $SL(2,\mathbbm{C})$ coordinate transformations on $\mathbbm{C}\mathbbm{P}^1$, leaving us with 18 physical moduli.} of 7-branes that can be varied.  Even though we have 24 7-branes, they are not perturbative 7-branes of a given string theory; rather each is a perturbative 7-brane of some $(p,q)$ theory.  Near each one we can use a perturbative description of the $(p,q)$ theory, but in going from one $(p,q)$ theory to another, one may double count states--there is no globally perturbative description in the generic case.  It turns out that one can take at most 16 7-branes to be $D$-branes of a given $(p,q)$ theory, say the $(1,0)$ theory.

This argument breaks down in the $N_f = 6,12$ cases since the field lines can run off to infinity.  Nonetheless, by examining the explicit solution, in say the $N_f = 6$ case, one can see that there are regions where $e^{-\Phi} \sim 1$.  One can have at most four ordinary $D7$-branes in a region of $e^{-\Phi} <<1$.  A better description in the region of the other two 7-branes is in terms of  $SL(2,\mathbbm{Z})$ transformed 7-branes \cite{Berg}.  Similarly, in the $N_f = 12$ case, one can have at most 8 ordinary $D7$-branes.

One can consider special configurations of 7-branes (in the compact or noncompact case) where the axidilaton is everywhere constant \cite{DM,Lerche}.  These correspond to taking certain combinations of $(p,q)$ branes coincident and can be classified by  orbifold limits of K3.

We will restrict attention to to those solutions that have a perturbative description in $g_s$.  In the \emph{constant} axidilaton case, with $N_f$ 7-branes, such solutions correspond to $N_f/6$ sets of 7-branes, each of which can be viewed as 4 ordinary $D7$-branes coincident with an $O7$-plane.  More precisely, the $O7$-plane is an $O7^-$ plane, with a charge of $-4$ relative to a $D7$-brane, and such that the gauge group on the $4D7+O7^{-}$ worldvolume is $SO(8)$.  $\tau$ is everywhere constant and can take on any value; thus there exists a $g_s \rightarrow 0$ limit.  Note, though, that even in this limit, the metric is nontrivial.  It can be made locally flat, but globally it still has a deficit angle.  The compact case is convincingly argued by Sen \cite{Sen} to be dual to Type I and heterotic with the gauge group broken to $SO(32) \rightarrow SO(8)^4$.

We will also consider deformations of these solutions obtained by pulling the 4 $D7$-branes off of the $O7$-plane.  These solutions have a varying axidilaton.  When this is done, the $O7$-plane splits nonperturbatively into two $(p,q)$ branes.  This was studied by Sen \cite{Sen}, who related the moduli space of the system to that of $\mathcal{N} = 2$ Seiberg-Witten theory with four flavors \cite{SeibergWitten}.  The splitting can also be seen from the supergravity solutions of \cite{Berg}.  However, the region of strong coupling remains localized and is shielded by the four $D7$-branes.  The string coupling remains small away from the system (and in the immediate vicinity of the $D7$-branes where $\tau_2 \rightarrow \infty$).  By encircling it at a safe distance one can measure its charge to be that of an $O7^-$ plane.  Around this region of an ``effective'' $O7$-plane, the supergravity solution has monodromy $S^2 = - \mathbbm{1}$.  $\tau$ is invariant under $S^2$.  For the $D$3-brane modes, though, this will play a crucial role in the analysis.

\subsection{Strings, symmetries, and supersymmetries}

We have four kinds of strings: closed and 7-7, 3-3, and 3-7 open strings.  The massless modes of the closed strings form the IIB supergravity multiplet in the bulk.  The massless 7-7 modes form an $\mathcal{N} = 1$, $d=8$ hypermultiplet, consisting of one complex scalar, one Weyl fermion, and the gauge field, all in the adjoint of the gauge group.  We denote these fields as $(\phi^{(D7)},\psi^{(D7)}, A_{\mu,\alpha}^{(D7)})$.  For simplicity in the following discussion, we will focus on one set of 4 $D7$'s $+$ $O7^-$.  Then if the 4 $D7$-branes are separated from the $O7$-plane, the gauge group is $U(4)_f$, while in the orientifold limit it is enhanced to $SO(8)_f$.

\subsubsection{3-3 strings, monodromy, and the color gauge group}

The story of the 3-3 strings is far more interesting.  The massless modes form the familiar content of $\mathcal{N} = 4$ Super-Yang-Mill's: three complex scalars, four Weyl fermions and the gauge field.  Following\footnote{Except that the fermions are in the $\mathbf{4}$ of $SU(4)_R$ and the supercharges are in the $\bar{\mathbf{4}}$ in our conventions.} the notation of \cite{Sohnius}, we denote the field content as
\begin{eqnarray} (M^{ij}, \psi^i, A_m). \end{eqnarray}
Here the superscript $i,j = 1,\ldots,4$ is an index in the $\mathbf{4}$ of $SU(4)_R$ while subscript $i,j$ is an index in the $\bar{\mathbf{4}}$.  The three complex scalars have been packaged into an antisymmetric matrix $M^{ij} = -M^{ji}$ that additionally has a reality constraint $(M^{ij})^\dag = \frac{1}{2} \epsilon_{ijkl} M^{kl} \equiv M_{ij}$.  These fields are adjoint valued, but what is the gauge group?

Normally one would expect the gauge group to be $U(N_c)$ for $N_c$ stacked branes, but what is the effect of the orientifold plane?  Around the orientifold plane there is an $SL(2,\mathbbm{Z})$ monodromy $S^2 = -\mathbbm{1}$.  This corresponds to the transformation $\Omega (-1)^{F_L}$ on closed string modes of the IIB theory, where $\Omega$ is worldsheet orientation reversal and $(-1)^{F_L}$ flips the sign of all Ramond states on the left \cite{Sen}.  For the 3-3 strings it corresponds to $\Omega$, which, for the massless modes amounts to complex conjugation of the $U(N_c)$ representation and a $\mathbbm{Z}_2$ action on the string wavefunction.  For strings with Neumann boundary conditions the $\mathbbm{Z}_2$ action is $-1$, while for strings with Dirichlet boundary conditions it is $+1$.  Hence, $\Omega$ gives a $-1$ for the gauge field and a $+1$ for the scalars and fermions in addition to exchanging the Chan-Paton indices.   Modes that propagate around the orientifold plane only come back to themselves up to this transformation.  It is a discrete $\mathbbm{Z}_2$ gauge symmetry and we would like to extend the $U(N_c)$ group to include it. This can be done using the following construction.\footnote{We thank G. Moore for explaining this to us.}

Automorphisms of the Lie algebra of $U(N_c)$ consist of $\mathbbm{Z}_{N_c}$ transformations which are elements of $U(N_c)$ and hence inner automorphisms, and a $\mathbbm{Z}_2$ element which acts as a reflection of the Dynkin diagram and is not an element of $U(N_c)$ and so is an outer automorphism. Using this $\mathbbm{Z}_2$ element, which we will call $\sigma$, one can construct a semi-direct product of $U(N_c)$ with $\mathbbm{Z}_2$, or a $\mathbbm{Z}_2$ extension of $U(N_c)$ which we will call $\overline{U(N_c)}$:
\begin{equation}
1 \rightarrow U(N_c) \rightarrow \overline{U(N_c)} \rightarrow \mathbbm{Z}_2 \rightarrow 1 \ .
\end{equation}
In terms of group elements, let $h_i$ denote elements of $U(N_c)$. We define a multiplication rule for elements $(h_i, \sigma)$ by
\begin{eqnarray}
(h_i,\sigma) (h_j,\sigma) & = & (h_i h_j^*,1), \\
(h_i, \sigma) (h_j, 1) & = & (h_i h_j^*,\sigma), \\
(h_i, 1) (h_j,\sigma) & = & (h_i h_j, \sigma).
\end{eqnarray}
One can easily check that this defines a group which we call $\overline{U(N_c)}$.

Since we are going to consider string defects (the $D3/O7$ intersections) in $\mathcal{N} = 4$ SYM's with holonomy corresponding to $\sigma \equiv \Omega$, we, according to the philosophy of discrete gauge symmetry, declare that our gauge group is $H = \overline{U(N_c)}$.  Furthermore, since $\sigma$ takes representations to their complex conjugates, we can view it as a generalized charge conjugation operator of the type studied in \cite{Slansky}, and therefore strings with $\sigma$ holonomy should be thought of as Alice strings \cite{Schwarz,BLeeP,BLoP}.

In the presence of solitons there can be obstructions to the global extension of the local symmetry group $H$ so that only some subgroup $\tilde H \subset H$ is the group of globally well-defined symmetries \cite{Schwarz,NelsonManohar,BMMNSZ,NelsonColeman}.  Following the discussion in Sec. 2 of \cite{Alford}, a string in a theory with unbroken gauge group $H$ and holonomy $U(2 \pi)$ has a globally defined subgroup $\tilde H$ which is the centralizer of $U(2 \pi)$ in $H$. For the strings considered here with $U(2 \pi)=\sigma$ and $H= \overline{U(N_c)}$ this subgroup is the set of $(h,s) \in H$  with $s={1,\sigma}$ obeying
\begin{equation}
(1,\sigma) (h,s) = (h,s) (1,\sigma)
\end{equation}
which implies that $h=h^*$. Since $h \in U(N_c)$, this implies that $h^T = h^{-1}$, that is that $h \in O(N_c)$. Thus the globally defined group is $\tilde H = O(N_c) \times \mathbbm{Z}_2$ (the product is now direct since $\sigma$ acts trivially on $h \in O(N_c)$).  Under this subgroup, the adjoint of $U(N_c)$ decomposes into the symmetric plus antisymmetric tensor representation of $O(N_c)$.

\subsubsection{3-7 strings and preserved supersymmetry}

Consider the $D7/D3$ intersection.  The number of Dirichlet-Neumann plus Neumann-Dirichlet directions is eight.  The system is supersymmetric, preserving 1/4 of the IIB supercharges; however, the NS zero-point energy is $1/2$ so there are only Ramond ground states.  After imposing the GSO projection, these form a single Weyl spinor of $Spin(1,1)$, transforming in the $(\mathbf{N}_c, \bar{\mathbf{4}})$ of $O(N_c) \times U(4)_f$ (or the $(\mathbf{N}_c, \mathbf{8})$ of $O(N_c) \times SO(8)_f$).  How can supersymmetry be preserved if the massless spectrum on the intersection has only fermionic degrees of freedom?  The fermion must be a singlet under the supersymmetry action.  We will take the fermion to be left-handed, denoting it $q_L$.  Then from the $1+1$-dimensional point of view, the supersymmetry is $\mathcal{N} = (0,8)$; all of the supercharges are right-handed.  We denote them by $Q_{Ri}, Q_{R}^{\dag i}$.

Finally, the ten-dimensional local Lorentz symmetry of IIB is broken down to $SO(1,1) \times SO(2) \times SO(6)$ by this setup.  The $SO(2)$, corresponding to rotations in the plane transverse to the 7-branes, is only preserved in the maximally symmetric case where all $D7$-branes and $O7$-planes coincide.  The transformation properties of the 3-3 modes, 3-7 modes, and supercharges under the full symmetry group are listed in Table \ref{table1}.

\TABLE{ \label{table1}
\begin{tabular}{l|l|l}
& $SO(1,1) \times SO(2) \times SO(6)$ & $O(N_c) \times U(4)_f(SO(8)_f)$ \\
\hline
$M^{ij}$ & $(0,0,\mathbf{6})$ & $(\mathbf{N}_{c}^2,\mathbf{1})$ \\
\hline
$\psi^i$ & $(\frac{1}{2} , \frac{1}{2}, \mathbf{4}) + (-\frac{1}{2}, - \frac{1}{2}, \mathbf{4})$ & $(\mathbf{N}_{c}^{2},\mathbf{1})$  \\
\hline
$A_\mu, A_{z,\bar{z}}$ & $(1,0,\mathbf{1}),(0, \pm1, \mathbf{1})$ & $(\mathbf{N}_{c}^2,\mathbf{1})$  \\
\hline
$q_L$ & $(\frac{1}{2}, 0, \mathbf{1})$ & $(\mathbf{N}_c,\bar{\mathbf{4}}(\mathbf{8}))$\\
\hline
$Q_{Ri}, Q_{R}^{\dag i}$ & $(-\frac{1}{2}, -\frac{1}{2}, \bar{\mathbf{4}}), (-\frac{1}{2}, \frac{1}{2}, \mathbf{4})$ & $(\mathbf{1},\mathbf{1})$
\end{tabular}
\caption{Transformation properties of $D3$-brane and intersection massless modes and preserved supercharges.  We specify the charges of the fields under the action of the Abelian groups and the dimensions of their representations for non-Abelian groups.}}

\section{Anomalies, the effective action, and $D3$-brane zero-modes}

Having described the system we will be studying we would like to identify the zero-modes and
hence the low-energy effective action describing massless excitations on the string intersection.
We first present a puzzle, and then resolve it later using both an index theory calculation and an
explicit construction of zero-modes.

\subsection{An anomaly puzzle} \label{lackof}

Let us first suppose that the four $D7$-branes are separated from the $O7$-plane.  The chiral fermions, $q_L$, localized at the $D3/D7$ intersection have both gauge and gravitational anomalies.  The well established mechanism of anomaly inflow from the $D3$- and $D7$-branes cancels this zero-mode anomaly \cite{Harvey1,CheungYin,Minasian}; we will not review the details here.  However, consider the $D3/O7$ intersection.  There are (apparently) no zero-modes localized at the intersection--at least not from open string quantization.  (Note also, this is an orientifold, not an orbifold--there are no twisted closed string sectors).  On the other hand, there is inflow.

The anomalous couplings on the $D3$-branes and $O7$-plane \cite{Dasgupta,Stefanski,ScruccaSerone} are given by:
\begin{eqnarray}  S_{D3}^{\textrm{WZ}} & = & \frac{\mu_3}{2} \int_{\Sigma^{4}} N_c C_4 - G \wedge Y(D3)^{(0)} \\
S_{O7}^{\textrm{WZ}} & = & \frac{\mu_7'}{2} \int_{\Sigma^{8}} C_8 - G \wedge Y(O7)^{(0)} \ , \end{eqnarray}
where $G$ is the sum of R-R form field strengths and the $Y$ are characteristic polynomials
\begin{eqnarray} Y(D3) & = & ch(F_c) \wedge \sqrt{ \frac{ \hat{A}(R_{T\Sigma^4}) }{ \hat{A} (R_{N\Sigma^4}) } } \ , \\
Y(O7) & = & \sqrt{ \frac{\hat{L}(R_{T\Sigma^8}/4)}{ \hat{L}( R_{N\Sigma^8}/4)} } \ . \label{O7anomalouscouplings}\end{eqnarray}
We are using standard conventions in the anomaly literature, where $4\pi^2 \alpha' = 1$.  This sets the IIB $Dp$-brane charge and the constant in front of the supergravity action to $\mu_p = (2\kappa_{10}^2)^{-1} = 2\pi$.  The $Op$-plane charge is given by $\mu_p' = - 2^{p-5} \mu_p$.  We are also using the descent notation for characteristic classes: $Y - Y_0 = dY^{(0)}$, where $Y_0$ is the constant piece, and under a gauge transformation $\delta Y^{(0)} = d Y^{(1)}$.  We write Chern classes without a subscript if they are evaluated in the fundamental representation, and will denote the dimension of the representation in the subscript otherwise.  $\hat{L}$ is the Hirzebruch L-polynomial and $\hat{A}$ the A-roof genus.  They can be expanded in Pontryagin classes as
\begin{eqnarray} \hat{A}(R) & = & 1 - \frac{1}{24} p_1(R) + \cdots \ , \nonumber \\
\hat{L}(R/4) & = & 1 + \frac{1}{3 \cdot 4^2} p_1(R) + \cdots \ . \end{eqnarray}
Following the standard anomaly inflow analysis, the gauge variation of the action is given by $2\pi$ times the integral over the intersection of the descent of an anomaly polynomial: $\delta S^{\textrm{WZ}} = 2\pi \int I^{(1)}$, where
\begin{eqnarray} I^{inf.D3/O7} & = & - (2\kappa_{10}^2) \frac{\mu_3 \mu_7'}{2 (2\pi)} ( Y(D3) \tilde{Y}(O7) + Y(O7) \tilde{Y}(D3) ) \nonumber \\
& = & 2 ( Y(D3) \tilde{Y}(O7) + Y(O7) \tilde{Y}(D3) ).  \end{eqnarray}
Here, $\tilde{Y}$ is given by conjugating the representation of the gauge group in $Y$, but this will not make a difference in our case.  Since our intersection is two-dimensional, we are interested in the 4-form part
\begin{eqnarray}   I_{4}^{inf.D3/O7} & = & 4 c_2(F_c) + 4 N_c \displaystyle\biggl( - \frac{1}{48} p_1(R_{T\Sigma^4}) + \frac{1}{48} p_1(R_{N\Sigma^4}) +  \nonumber \\
& &  \qquad \qquad \qquad \quad + \frac{1}{96} p_1(R_{T\Sigma^8}) - \frac{1}{96} p_1(R_{N\Sigma^8}) \displaystyle\biggr).\end{eqnarray}
Let us decompose the ten-dimensional tangent bundle according to $TM = T_{SO(1,1)} \oplus N_{SO(6)} \oplus \tilde{N}_{SO(2)}$, and use $p_1(E \oplus F) = p_1(E) + p_1(F)$ to arrive at
\begin{eqnarray} \label{D3O7Inflow} I_{4}^{inf.D3/O7} & = & 4 c_2(F_c) - \frac{N_c}{24} p_1(T_{SO(1,1)}) + \frac{N_c}{8} p_1(N_{SO(6)}) - \frac{N_c}{8} p_1(\tilde{N}_{SO(2)}). \end{eqnarray}

One may raise objections to this analysis.  As we reviewed above, Sen \cite{Sen} has shown that when separating the $D7$-branes from the $O7$-plane, the region near the $O7$-plane has $e^{-\Phi} \sim 1$.  Nonperturbative effects cause the $O7$-plane to be split into two $SL(2,\mathbbm{Z})$ transformed 7-branes.  Nonetheless, if we circle around the region of strong coupling at a safe distance, where a perturbative description is valid, it ``looks'' just like an $O7$-plane--it has the same charge and induces the same $SL(2,\mathbbm{Z})$ monodromy on the spectrum.  Furthermore, one does not expect well-behaved corrections (perturbative or not) to alter the result of an anomaly computation.

We can further demonstrate that there is something missing by considering the question of anomaly inflow and cancellation in the orientifold limit, where the $D$-branes and $O$-plane coincide.  In this limit, the string coupling is constant and can be taken arbitrarily small (to zero, in fact!).  The inflow onto the intersection is now a sum of inflows from the $D3/O7$ intersection and the $D3/D7$ intersection.  The $D3/O7$ inflow is given above \eqref{D3O7Inflow}.  We now have an $SO(8)$ gauge group on the $D7$-branes; there are eight $D7$-branes counting their images.  Note, though, with this counting, we must use the Type I charge $\mu_{p}^{(\textrm{I})} = \frac{1}{2} \mu_{p}^{(\textrm{II})}$.  We thus find
\begin{eqnarray} I^{inf. D3/4D7} & = &  -\frac{1}{4} (Y(D3) \tilde{Y}(D7) + Y(D7) \tilde{Y}(D3) ), \end{eqnarray}
with
\begin{eqnarray} Y(D7) & = & ch(F_f) \wedge \sqrt{ \frac{ \hat{A}(R_{T\Sigma^8}) }{ \hat{A} (R_{N\Sigma^8}) } } \ , \label{D7anomalouscouplings} \end{eqnarray}
and $Y(D3)$ as before.  Note the Chern form in $Y(D7)$ is being evaluated in the $\mathbf{8}_v$ of $SO(8)$.  We find the 4-form component of this to be
\begin{eqnarray} I_{4}^{inf.D3/4D7} = -4 c_2(F_c) - \frac{N_c}{2} c_2(F_f) + \frac{N_c}{6} p_1 (T_{SO(1,1)}) \end{eqnarray}
so that the sum of inflows onto the orientifold intersection is given by
\begin{eqnarray} I_{4}^{inf.D3/4D7} + I_{4}^{inf.D3/O7} & = & - \frac{N_c}{2} c_2(F_f) + \frac{N_c}{8} (p_1(T_{SO(1,1)})  \nonumber \\
& & + p_1(N_{SO(6)}) - p_1(\tilde{N}_{SO(2)})). \end{eqnarray}
Observe that the $c_2(F_c)$ anomaly cancels between the the $D3/O7$ and $D3/D7$ inflows.  This makes physical sense, of course.  This term in the anomaly comes from two places: the coupling of $\textrm{tr}(F_{c}^2)^{(0)}$ to $G_1$ in the $D3$-brane action, and the gauge variation of $C_8$ in the $D7$-brane and $O7$-plane action.  But in the orientifold limit, $G_1$, or equivalently $G_9$, is no longer sourced because the R-R charge cancels locally between the $D7$-branes and $O7$-plane.  Thus, there is no $C_8$ one-point coupling in $S_{D7}^{\textrm{WZ}} + S_{O7}^{\textrm{WZ}}$, and so $dG_1 = 0$.

There is also a contribution to the anomaly from the 3-7 strings localized on the intersection.  These are left-handed Weyl fermions.  The $q_L$ transform in the $(\mathbf{N}_c,\mathbf{8})$ of $O(N_c) \times SO(8)$ as do the $q_{L}^\dag$.  However, in the orientifold limit, the $q_L$ and $q_{L}^\dag$ are related by the orientifold projection.  Worldsheet orientation reversal sends $q_L$ to $q_{L}^\dag$, so only the linear combination $q_L + q_{L}^\dag$ survives.\footnote{This 1/2 can also be understood as follows.  Moving the $D7$-branes to the $O7$-plane should not produce any new massless 3-7 modes. (There are no stretched strings to become massless as in the 7-7 case).  Before moving the $D7$'s to the $O7$ we had $2 \cdot 4N_c$ 3-7 states.  Now we still have $8N_c$ states.}  This produces an extra factor of $1/2$ in the index formula for the zero-mode anomaly: $\delta S^{z.m.} = 2\pi \int (I^{z.m.})^{(1)}$, with
\begin{eqnarray} I_{4}^{q} & = & \frac{1}{2} ch_{(\mathbf{N}_c,\mathbf{8})} (F_c \oplus F_f) \wedge \hat{A}(T_{SO(1,1)}) |_4 \nonumber \\
& = & 4 c_2(F_c) + \frac{N_c}{2} c_2(F_f) - \frac{N_c}{6} p_1(T_{SO(1,1)}).  \end{eqnarray}

As expected, the zero-mode anomaly precisely cancels the $D3/D7$ inflow anomaly, and the $D3/O7$ inflow remains uncancelled as before:
\begin{eqnarray} I_{4}^{q} + I_{4}^{inf.D3/4D7} + I_{4}^{inf.D3/O7} & = & I_{4}^{inf.D3/O7} \ . \end{eqnarray}
Clearly something is missing.  As we will see, the correct resolution of this puzzle will depend on the value of $N_f$.  First, however, let us give a more detailed description of this system that will be useful in the following.

\subsection{Low energy effective action} \label{Section3.2}

In this section we would like to write down an effective action governing fluctuations of the system about the supergravity background discussed above.  We will work in the low-energy limit  $\alpha' \rightarrow 0$.  In this limit the fluctuations of bulk and 7-brane modes decouple from the rest of the system, and we will not consider them further.

The action for the 3-7 strings is\footnote{We assume that, in the compact case, there are no Wilson lines turned on.}
\begin{eqnarray} S_{\textrm{3-7}} & = & \frac{1}{2\pi} \int_{I} d^2 x \bar{q} \gamma_{(2)}^\mu (i \partial_\mu - A_\mu + A_{\mu}^{D7}) L_{(2)} q \ . \end{eqnarray}
The $q$ have indices in the bi-fundamental of $O(N_c) \times U(4)_f$ (or $O(N_c) \times SO(8)_f$) that are suppressed.  After canonically normalizing the gauge fields one sees that the coupling to $A_{\mu}^{D7}$ vanishes as $\alpha' \rightarrow 0$ and the flavor group becomes a global symmetry.  For the $SO(1,1)$ gamma matrices we take $\gamma_{(2)}^0 = i \sigma^2$ and $\gamma_{(2)}^1 = \sigma^1$, so that $\bar{\gamma}_{(2)} = \sigma^3$, where the $\sigma^i$ are Pauli matrices.  Then $L_{(2)} q = \frac{1}{2} (1 + \sigma^3) q = (q_L,0)^T$, so that
\begin{eqnarray} S_{\textrm{3-7}} & \rightarrow & \frac{1}{2\pi} \int_{I} d^2 x q_{L}^\dag (i \partial_- - A_-) q_L \ , \label{37action} \end{eqnarray}
where $\partial_- = \frac{1}{2}(\partial_0 - \partial_1)$ etc.  Thus, in our conventions, left-handed corresponds to left-moving in $1+1$ dimensions, where by ``left-moving'' we really mean $q_L = q_L(x^+)$.  Using the fact that the $q_L$ are supersymmetry singlets, one can verify that, in standard Wess and Bagger notation, the subset of preserved $\mathcal{N} = 4$, $d=4$ supercharges is $Q_{2}^j, \bar{Q}_{\dot{2} j}$, $j = 1,\ldots,4$.  These have the anticommutation relations
\begin{eqnarray} \{ Q_{2}^j, \bar{Q}_{\dot{2}k} \} & = & 2 P_- \delta^{j}_{k} \ , \end{eqnarray}
ie. they are right-handed.

Next consider the 3-3 strings.  We can obtain the low energy effective action by expanding $S = S_{\textrm{DBI}} + S_{\textrm{WZ}}$ in the background of nontrivial metric, dilaton, and axion.  The bosonic part can be obtained by starting from the non-Abelian action given in \cite{Myers}, converting to Einstein frame, and keeping $O(\alpha'^0)$ terms.  This procedure is straightforward and we will simply give the result.  After converting to our notation we find
\begin{eqnarray} S_{\textrm{3-3}}^{bos.} & = & - \frac{1}{2\pi} \int d^4 x \sqrt{-g} \textrm{tr} \displaystyle\biggl( \frac{1}{4} e^{-\Phi} F_{mn} F^{mn} + \frac{1}{8} C_0 \epsilon^{mnpq} F_{mn} F_{pq} + \nonumber \\
& &  + \frac{1}{2} \mathcal{D}_m M_{ij} \mathcal{D}^m M^{ij} - \frac{1}{4} e^\Phi [ M_{ij} , M_{kl} ] [ M^{ij}, M^{kj} ] \displaystyle\biggr), \label{33bosonic} \end{eqnarray}
where $\epsilon^{0123} = (-g)^{-1/2}$.  $\mathcal{D}_m$ is the spacetime and gauge covariant derivative; acting on scalars it is simply
\begin{eqnarray} \mathcal{D}_m M^{ij} = \partial_m M^{ij} + i[ A_m, M^{ij}]. \end{eqnarray}
One could rescale the scalar fields by $M^{ij} \rightarrow e^{-\Phi/2} M^{ij}$ and identify $g_{YM}^2 \equiv 2\pi e^\Phi$.  Then part of the action will have the standard $\mathcal{N} = 4$ form, in a curved spacetime and with a spacetime dependent $g_{YM}^2$ and $\theta$-angle.  However, the derivative on the scalars will also generate a coupling to $\partial_m \Phi$ if the dilaton is nonconstant.  Thus the effective action is not quite what one might guess by naively generalizing the usual $\mathcal{N} = 4$ action.

The fermionic $Dp$-brane actions in general bosonic supergravity backgrounds have been worked out very explicitly to quadratic order in fermions in \cite{MMS}.  They restricted attention to the Abelian case, but to quadratic order in 3-3 modes, one can trivially generalize to the non-Abelian case by adding a trace.  In other words, their results are applicable up to three-point couplings of the fermions with the gauge field and the scalars.  After some work, one reduces the general formula of \cite{MMS}, in the case of $D3$-branes in a background metric, dilaton, and R-R 1-form field strength, to the following effective action:
\begin{eqnarray} \label{33fermionaction}  S_{\textrm{3-3}}^{ferm.} & = &  \frac{i}{2\pi} \int d^4 x \tau_2 \sqrt{-g}  \textrm{tr} \displaystyle\biggl( \bar{\psi}_i \gamma^m ( D_m + \frac{i}{2} Q_m ) L \psi^i \displaystyle\biggr) +  \nonumber \\
& & + O(\psi^2 A, \psi^2 M). \end{eqnarray}
This is an important result, and we give the details of the calculation in Appendix \ref{AppendixA}.  There are a couple of comments to be made.  $D_m$ is the spacetime covariant derivative,
\begin{eqnarray} D_m & = & \partial_m + \frac{1}{4} \omega_{ab,m} \gamma^{ab} \ , \end{eqnarray}
where $\omega_{ab,m}$ is the usual spin connection associated with the metric.  $Q_m$ is defined by
\begin{eqnarray} Q_m \equiv \frac{ \partial_m (\tau + \bar{\tau})}{2i (\tau - \bar{\tau})} \ . \end{eqnarray}
This is the pullback over the spacetime of the Kahler connection $Q = \frac{1}{2i} (\partial_{\tau} \mathcal{K} d\tau - \partial_{\bar{\tau}} \mathcal{K} d\bar{\tau})$ on the special Kahler manifold $SL(2,\mathbbm{R})/U(1)$, where $\mathcal{K} = \log{\tau_2}$ is the Kahler potential \cite{Berg}.
The coupling of the fermions to this connection will play a crucial role in the following.

Later, we will also require the full fermionic action, including the three-point couplings to the gauge field and the scalars.  These couplings can be deduced by using gauge invariance and requiring that the action reduce to the standard $\mathcal{N}=4$ case when the axidilaton is constant.  These considerations lead us to
\begin{eqnarray} S_{\textrm{3-3}}^{ferm.} & = & \frac{i}{2\pi} \int d^4 x  \sqrt{-g} \textrm{tr} \displaystyle\biggl\{ \tau_2 \displaystyle\biggl( \bar{\psi}_i \gamma^m ( \mathcal{D}_m + \frac{i}{2} Q_m ) L \psi^i \displaystyle\biggr) +  \nonumber \\
& & +  \sqrt{\tau_2} \displaystyle\biggl( (L\psi)^i [ (L\psi)^j, M_{ij}] + (\bar{\psi} R)_i [ (\bar{\psi} R)_j, M^{ij}] \displaystyle\biggr) \displaystyle\biggr\}, \label{33fermionic} \end{eqnarray}
where $\mathcal{D}_m = D_m + i[A_m, \quad]$.

\subsection{Zero-modes on the $D3$-brane} \label{Section3}

IIB string theory is believed to be a consistent, anomaly free theory.  Therefore the anomaly puzzle indicates that $(1)$ we don't have the proper anomalous couplings for this type of $D$-brane/$O$-plane intersection to give the correct anomaly inflow, \emph{and/or} $(2)$ we haven't accounted for all of the massless, chiral zero-modes localized on the intersection.  Let us first investigate the latter possibility.

\subsubsection{The index calculation}

If there are more zero-modes, then they must come from the 3-3 strings.  We have already accounted for the 3-7 strings, and the 7-7 and closed strings decouple, becoming free in the $\alpha' \rightarrow 0$ limit (while the anomaly inflow to the $D3/O7$ intersection does not vanish in this limit).  Let us analyze the effective action \eqref{33fermionaction} with this in mind.

Since the supergravity background depends only on the transverse coordinates $z,\bar{z}$, it will be convenient to work with gamma matrices $\gamma^a = (\gamma^{\underline{\mu}}, \gamma^{\underline{z}}, \gamma^{\underline{\bar{z}}} )$ satisfying
\begin{eqnarray} \{ \gamma^a, \gamma^b \} = 2 \left( \begin{array}{c c c} \eta^{\mu\nu} & & \cr & 0 & 2 \cr & 2 & 0 \end{array} \right)^{ab} . \end{eqnarray}
{}From the metric $ds^2 = \eta_{\mu\nu} dx^\mu dx^\nu + e^{a(z,\bar{z})} dz d\bar{z}$, one deduces the vielbeins
\begin{eqnarray} e^{\underline{\mu}} = dx^\mu \ , \qquad e^{\underline{z}} = e^{a/2} dz \ , \qquad e^{\underline{\bar{z}}} = e^{a/2} d\bar{z} \ . \end{eqnarray}
(Note that this implies $e_{\underline{z}} = \frac{1}{2} e^{a/2} d\bar{z}, e_{\underline{\bar{z}}} = \frac{1}{2} e^{a/2} dz$).  From these it is straightforward to obtain
\begin{eqnarray} \gamma^m \omega_{ab,m} \gamma^{ab} & = & e^{-a/2} ( \gamma^{\underline{z}} \partial a + \gamma^{\underline{\bar{z}}} \bar{\partial} a ), \end{eqnarray}
where we are using the shorthand $\partial \equiv \partial_{z}, \bar{\partial} \equiv \partial_{\bar{z}}$.

Let us choose a basis for the gamma matrices that is convenient for the decomposition $SO(1,3) \rightarrow SO(1,1) \times SO(2)$:
\begin{eqnarray} \gamma^\mu = \mathbbm{1}_2 \otimes \gamma_{(2)}^\mu \ , \qquad \gamma^{\underline{2},\underline{3}} = \sigma^{1,2} \otimes \bar{\gamma}_{(2)} \ . \end{eqnarray}
Note in particular that
\begin{eqnarray} \gamma^{\underline{z}} \equiv \sigma^{\underline{z}} \otimes \bar{\gamma}_{(2)} = \left(\begin{array}{c c} 0 & 2 \bar{\gamma}_{(2)} \\ 0 & 0 \end{array}\right), \qquad \gamma^{\underline{\bar{z}}} \equiv \sigma^{\underline{\bar{z}}} \otimes \bar{\gamma}_{(2)} = \left(\begin{array}{c c} 0 & 0 \\ 2 \bar{\gamma}_{(2)} & 0 \end{array}\right). \end{eqnarray}
Then the fermionic action becomes
\begin{eqnarray} S_{\textrm{3-3}}^{ferm.} & = & \frac{i}{2\pi} \int d^4 x \tau_2 \sqrt{-g} \textrm{tr} \displaystyle\biggl( \bar{\psi}_i  \mathcal{O}_{Dirac} L \psi^i \displaystyle\biggr), \qquad \textrm{with} \\
\mathcal{O}_{Dirac} & = & \left(\begin{array}{c c} \gamma_{(2)}^{\mu} \partial_\mu & 2 e^{-a/2} (D_z + \frac{i}{2} Q_z) \bar{\gamma}_{(2)} \\ 2 e^{-a/2} (D_{\bar{z}} + \frac{i}{2} Q_{\bar{z}} ) \bar{\gamma}_{(2)} & \gamma_{(2)}^\mu \partial_\mu \end{array}\right).  \end{eqnarray}

Now choose the $\gamma_{(2)}^\mu$ as before:
\begin{eqnarray} \gamma_{(2)}^0 = i \sigma^2 \ , \qquad \gamma_{(2)}^1 = \sigma^1 \qquad \Rightarrow \qquad \bar{\gamma}_{(2)} = \sigma^3 \ . \end{eqnarray}
Then we have
\begin{eqnarray} L = \left( \begin{array}{c c} L_{(2)} & 0 \\ 0 & R_{(2)} \end{array}\right), \qquad \textrm{where} \quad L_{(2)} = \left(\begin{array}{c c} 1 & 0 \\ 0 & 0 \end{array}\right), \qquad R_{(2)} = \left( \begin{array}{c c} 0 & 0 \\ 0 & 1 \end{array}\right). \end{eqnarray}
Therefore we make the ansatz
\begin{eqnarray} L \psi^i = \left(\begin{array}{c} \lambda_{L+}^i(x^\mu, z, \bar{z}) \\ \lambda_{R-}^i(x^\mu, z, \bar{z}) \end{array}\right) = \left( \begin{array}{c} \xi_{L}^i(x^\mu) f_+(z,\bar{z}) \\ 0 \\ 0 \\ \xi_{R}^i(x^\mu) f_-(z,\bar{z}) \end{array}\right). \end{eqnarray}
The $\xi$ are one-component complex Weyl fermions in $d = 1+1$ dimensions, while the $\pm$ indicates the sign of the $SO(2)$ charge, consistent with the decomposition $\mathbf{2} \rightarrow ( \frac{1}{2} )_+ + ( - \frac{1}{2} )_-$ under $SO(1,3) \rightarrow SO(1,1) \times SO(2)$.  Plugging all of this in, we are left with the action
\begin{eqnarray} S_{\textrm{3-3}}^{ferm.} & = &  -\frac{i}{2 \pi} \int d^4 x \tau_2 \sqrt{-g} \textrm{tr} \displaystyle\biggl( \xi_{Li}^\dag \partial_- \xi_{L}^i | f_+ |^2 + \xi_{Li}^\dag \xi_{R}^i e^{-a/2} f_{+}^\ast (D_z + \frac{i}{2} Q_z) f_- \nonumber \\
& & + \xi_{Ri}^\dag \xi_{L}^i e^{-a/2} f_{-}^\ast (D_{\bar{z}} + \frac{i}{2} Q_{\bar{z}}) f_+ + \xi_{Ri}^\dag \partial_+ \xi_{R}^i |f_-|^2 \displaystyle\biggr). \end{eqnarray}

Thus we see that the number of left-moving, massless zero-modes is equal to the number of linearly independent, normalizable solutions to $(D_{\bar{z}} + \frac{i}{2} Q_{\bar{z}}) f_+ = 0$, while the number of right-moving zero-modes equals the number of normalizable solutions to $(D_z + \frac{i}{2} Q_z) f_- = 0$.  We can construct the two-dimensional Euclidian Dirac operator and spinor:
\begin{eqnarray} \slashed{\mathcal{D}}_{Q} \equiv \sigma^z (D_z + \frac{i}{2} Q_z) + \sigma^{\bar{z}} (D_{\bar{z}} + \frac{i}{2} Q_{\bar{z}}), \qquad f \equiv \left( \begin{array}{c} f_+ \\ f_- \end{array}\right). \end{eqnarray}
Then the number of left-moving modes (solutions of $\slashed{\mathcal{D}}_{Q} L f = 0$) minus the number of right-moving modes (solutions of $\slashed{\mathcal{D}}_{Q} R f = 0$) is seen to be equal to the index

\begin{eqnarray} \label{indextheorem} \textrm{ind} (i \slashed{\mathcal{D}}_Q )_{(1/2)} = \frac{i}{2} \int_{\mathbbm{C} \mathbbm{P}^1} ch_{1} \mathcal{F}_Q = -\frac{1}{4\pi} \int_{\mathbbm{C} \mathbbm{P}^1} (\mathcal{F}_Q)_{z\bar{z}} dz d\bar{z} \ , \end{eqnarray}
where $(\mathcal{F}_Q)_{z\bar{z}} = \partial_z Q_{\bar{z}} - \partial_{\bar{z}} Q_z$.  But this integral can be easily evaluated.  Recall that the supergravity equations of motion imply that $\tau$ is an \emph{anti-holomorphic} function, $\tau = \tau(\bar{z})$.  (See Footnote \ref{footnote1} for why we must take $\tau$ anti-holomorphic instead of holomorphic).  Then one has
\begin{eqnarray} Q_z & = & \frac{\partial (\tau + \bar{\tau})}{2 i (\tau - \bar{\tau})} = - \frac{ \partial ( \tau - \bar{\tau})}{ 2 i (\tau - \bar{\tau})} = - \frac{1}{2i} \partial \log{\tau_2} \ , \\
Q_{\bar{z}} & = & \frac{ \bar{\partial} (\tau + \bar{\tau})}{2 i( \tau - \bar{\tau})} = \frac{\bar{\partial} ( \tau - \bar{\tau})}{ 2i (\tau - \bar{\tau})} = \frac{1}{2i} \bar{\partial} \log{\tau_2} \ , \end{eqnarray}
and so
\begin{eqnarray} (\mathcal{F}_Q)_{z\bar{z}} = \frac{1}{i} \partial \bar{\partial} \log{\tau_2} \ . \end{eqnarray}
Now, this is proportional to the volume element on the fundamental domain.  Since it is a modular invariant, we can pull the integral over the complex plane back to an integral over $F_0$.  Recalling that $\tau(\bar{z}) = \bar{j}^{-1}(\bar{f}(\bar{z}))$ wraps the fundamental domain $N_f$ times, we have
\begin{eqnarray}   \textrm{ind} (i \slashed{\mathcal{D}}_Q )_{(1/2)} & = &  \frac{i}{4\pi} \int_{\mathbbm{C}} \frac{\bar{\partial} \tau \partial \bar{\tau}}{(\tau - \bar{\tau})^2} dz d\bar{z} = \frac{i N_f}{4\pi} \cdot \frac{i}{2} \int_{F_0} \frac{d\tau_1 d\tau_2}{\tau_{2}^2} \nonumber \\
& = & -\frac{N_f}{8\pi} \cdot \frac{\pi}{3} = -\frac{N_f}{24}. \end{eqnarray}
This looks fractional except for the compact case, where $N_f = 24$.  In fact, the form of the index theorem above \eqref{indextheorem} is only correct on compact manifolds.  For the noncompact cases there is a surface term that must be properly accounted for \cite{APSI,APSII,APSIII}.

The index theorem we just used assumes that the zero-modes have periodic boundary conditions: $f(e^{2\pi i} z, e^{-2\pi i} \bar{z}) = f(z,\bar{z})$.  However, recall that the 3-3 modes should have a $\mathbbm{Z}_2$ monodromy around the $O7$-planes, which acts by worldsheet orientation reversal.  In terms of the globally preserved gauge group $O(N_c)$, we have fermions in the symmetric plus antisymmetric tensor reps: $\mathbf{N}_{c}^2 = \frac{\mathbf{N}_c (\mathbf{N}_c + \mathbf{1})}{\mathbf{2}} + \frac{\mathbf{N}_c (\mathbf{N}_c - \mathbf{1})}{\mathbf{2}}$.  Thus there are two cases to consider.  The states in the symmetric tensor representation should have a periodic wave function, while the states in the antisymmetric tensor representation should have an antiperiodic wave function.  In this way the total state is single valued and thus well defined.

This last point is somewhat subtle and deserves further comment.  Away from the $O7$-plane the gauge group on the $D3$-branes is locally $U(N_c)$.  Thus one can, locally, distinguish between a typical fermionic fluctuation $\psi$ and its conjugate $\bar{\psi}$.  If we run $\psi$ around the $O7$-plane it will come back as $\bar{\psi}$, but the orientifold action does not locally constrain the space of states.  The fermionic zero-mode is different, however.  It is, in a sense, a global object; we will eventually solve for the explicit wavefunction $f(z,\bar{z})$ and see that it is extended around the entire $O7$-plane.  As we go around the $O7$-plane the state must be single valued and, therefore, must be invariant under the orientifold action.  As a consequence, the zero-modes $\xi_L,\xi_R$ should only take values in $O(N_c)$.

We have so far found that there are $N_f/24$ right-handed zero-modes in the symmetric tensor representation of $O(N_c)$.  Now, antiperiodic boundary conditions can be obtained from the periodic case by the singular gauge transformation $\psi_p \rightarrow \psi_a = e^{-i \theta/2} \psi_p$, where $\theta(z) = Arg(z)$ is the azimuthal coordinate.  (The sign was determined from the fact that the periodic zero-mode has $SO(2)$ charge $-1/2$).  Note, though, that we require antiperiodicity around \emph{all} of the $O7$-planes.  If there are $N_f/6$ $O7$-planes located at positions\footnote{Away from the orientifold limit, there are not really $O7$-planes located at specific points $z_{O7}^{(k)}$.  However, there are regions of strong coupling that can be made relatively small and thought of as composite $O7$-planes.  If we only consider loops that go entirely around these regions, they can be approximated as $O7$-planes at locations $z_{O7}^{(k)}$.} $z_{O7}^{(k)}$, then the appropriate gauge parameter is $\alpha = \frac{1}{2} ( \theta_1 + \cdots + \theta_{N_f/6})$, where $\theta_k = \theta(z - z_{O7}^{(k)})$.  This corresponds to the transformation $\frac{i}{2} Q \rightarrow \frac{i}{2} Q - \frac{i}{2} (d\theta_1 + \cdots + d\theta_{N_f/6})$.  This is a singular transformation because ``$d^2\theta$''$\neq 0$.  It gives a contribution to the field strength localized at the $z_{O7}^{(k)}$.  We have
\begin{eqnarray} \textrm{ind}_{antisym.} (i \slashed{\mathcal{D}}_Q )_{(1/2)} & = & \textrm{ind}_{sym} (i \slashed{\mathcal{D}}_Q )_{(1/2)} - \frac{i}{2\pi} \int_{\mathbbm{C}} \frac{i}{2}(d^2\theta_1 + \cdots d^2\theta_{N_f/6}) \nonumber \\
& = & \textrm{ind}_{sym} (i \slashed{\mathcal{D}}_Q )_{(1/2)} + \frac{1}{4\pi} (2\pi) \frac{N_f}{6} \nonumber \\
& = & \textrm{ind}_{sym} (i \slashed{\mathcal{D}}_Q )_{(1/2)} + \frac{N_f}{12} =  \frac{N_f}{24} \ . \end{eqnarray}
Thus, in the compact case, there is one left-handed zero-mode, in the antisymmetric tensor representation of $O(N_c)$.

We have been implicitly considering the generic case, where the $D7$-branes are separated from the $O7$-planes, so that the axidilaton is varying.  In the orientifold limit, where $\tau$ becomes constant, one might rashly conclude that the index vanishes.  However, as we bring the $D7$-branes towards the $O7$-plane, we are bringing a region where $\tau \rightarrow i \infty$ near a region where $\tau \sim O(1)$.  This corresponds to $\partial \bar{\tau} \rightarrow \infty$ in the vicinity of the branes while $\partial \bar{\tau} \rightarrow 0$ away from them.
Thus it appears that $\mathcal{F}_Q$ is becoming $\delta$-function localized in such a way that the index is preserved.  This argument is heuristic.  On the other hand, it is born out by an explicit perturbative string theory analysis of the zero-modes in the orientifold limit.

Let us now briefly review the orientifold limit of the $N_f = 24$ system and its well known dual descriptions, as it will be useful in the following.  Note also that we will be focusing on the compact case exclusively in the next few subsections, and we will return to the noncompact cases at the end of section \ref{Section3.5}.

\subsubsection{String dualities and the orientifold limit}

The orientifold limit corresponds to an orbifold limit of the $\mathbbm{C}\mathbbm{P}^1$ where it becomes equivalent to $T^2/\mathbbm{Z}_2$.  The easiest way to see the zero-modes is by doing $T$-duality along both directions of the torus.  Note that $T$-duality is a perturbative symmetry of string theory, and so maps zero-modes to zero-modes.  The $D3/D7$-$O7$ intersection is mapped to $N_c$ $D1$-strings in Type I on $\mathbbm{R}^{1,7} \times \tilde{T}^2$.  The 3-7 strings are mapped to 1-9 strings, while the $D3$-brane zero-modes are mapped to the massless excitations of 1-1 strings.

The $1+1$-dimensional theory of massless modes on the worldsheet of $N_c$ coincident $D1$-strings in Type I has been well studied in the context of Type I/heterotic duality \cite{PolchinskiWitten}, and heterotic matrix strings \cite{BSS,BM,Lowe,Rey}.  It is a theory with $(0,8)$ supersymmetry, $O(N_c)$ gauge group, $SO(8)_R$ R-symmetry group, and it is uniquely specified by the Yang-Mills coupling $g_{1}$.  The action is
\begin{eqnarray} S_{D1} & =  & -\frac{1}{2\pi} \int d^2 x \textrm{tr} \displaystyle\biggl( \frac{1}{4 g_{1}^2} F_{\mu\nu} F^{\mu\nu} + \frac{1}{2} (\mathcal{D}_\mu X^I)^2 - \frac{g_{1}^2}{4} [X^I, X^J]^2 \nonumber \\
& & + \frac{i}{g_{1}^2} \lambda_{L}^T C \mathcal{D}_{-} \lambda_L + i \theta_{R}^T C \mathcal{D}_{+} \theta_R + 2i \lambda_{L}^T C \sigma^I [\theta_R,  X^I] - i \chi_{L}^\dag \mathcal{D}_- \chi_L \displaystyle\biggr). \nonumber \\ \label{D1action} \end{eqnarray}
The transformation properties of the field content under the gauge and global symmetries are given in Table \ref{Table2}.  The supersymmetry variations can be found in \cite{BSS}.  We have $\mathcal{D}_\mu = \partial_\mu + i[  a_\mu, \quad]$, with $a_\mu$ a $1+1$-dimensional gauge field.  $(a_\mu, \lambda_L)$ is a gauge multiplet in the antisymmetric (adjoint) representation of $O(N_c)$ and $(X^I,\theta_R)$ is a hypermultiplet in the symmetric tensor representation.  The $\chi$ are the supersymmetry singlets coming from 1-9 strings.  $\lambda_L,\theta_R$ are self-conjugate spinors in the $\mathbf{8}_s$ and $\mathbf{8}_c$ of $SO(8)_R$ respectively; the subscripts refer to their worldsheet chirality.  $C$ is the charge conjugation matrix of $SO(8)$ acting on both Weyl representations, and the $\sigma^I$ are Clebsch-Gordan coefficients for $\mathbf{8}_s \times \mathbf{8}_c \rightarrow \mathbf{8}_v$, with $I$ an index in the $\mathbf{8}_v$.

In Table \ref{Table2} we also indicate the zero-modes in the $D3/D7$-$O7$ system that are $T$-dual to the Type I $D1$-string modes.  Note that the assignments $\lambda_L = (\xi_{L}^i,\xi_{Li}^\ast )$, $\theta_R = (\xi_{R}^i, \xi_{Ri}^\ast)$ are consistent with the decompositions $\mathbf{8}_s \rightarrow \mathbf{4}_+ + \bar{\mathbf{4}}_-$ and $\mathbf{8}_c \rightarrow \mathbf{4}_- + \bar{\mathbf{4}}_+$ under $SO(8) \rightarrow SO(6) \times SO(2)$.  Futhermore, since the $\xi$ only take values in $O(N_c)$ the gauge representation assignments of $\lambda_L,\theta_R$ make sense.  The relations $a_\mu \leftrightarrow A_\mu$ and $X^I \leftrightarrow (A_z,A_{\bar{z}}, M^{ij})$ are standard from $T$-duality.  They do indicate to us, though, that away from the orientifold limit we should expect these (nonchiral) bosonic zero-modes as well.

\TABLE{ \label{Table2}
\begin{tabular}{r|c|c|c|l}
 & $ \begin{array}{c} D1 \\ \textrm{modes} \end{array} $  &  $T$-dual modes & $ \begin{array}{c} SO(1,1) \\ \times SO(8)_R \end{array} $ & $O(N_c) \times SO(32)$ \\
\hline
 gauge & $ \begin{array}{c} a_\mu \\ \lambda_L \end{array} $ & $ \begin{array}{c} A_\mu \\ \xi_{L}^i \oplus \xi_{Li}^\ast \end{array} $ & $ \begin{array}{c} (1, \mathbf{1}) \\ (1/2, \mathbf{8}_s) \end{array} $ & $( \frac{\mathbf{N}_c ( \mathbf{N}_c - \mathbf{1})}{ \mathbf{2}} , \mathbf{1} )$ \\
\hline
 hyper & $ \begin{array}{c} X^I \\ \theta_R \end{array} $ & $ \begin{array}{c} (A_z,A_{\bar{z}}, M^{ij}) \\ \xi_{R}^i \oplus \xi_{Ri}^\ast \end{array} $ & $ \begin{array}{c} (0, \mathbf{8}_v) \\ (-1/2, \mathbf{8}_c) \end{array} $ & $( \frac{\mathbf{N}_c ( \mathbf{N}_c + \mathbf{1})}{ \mathbf{2}} , \mathbf{1} )$ \\
 \hline
 hyper & $\chi_L$ & $q_L$ & $(1/2, \mathbf{1})$ & $(\mathbf{N}_c, \mathbf{32})$
\end{tabular}
\caption{Field content of Type I $D1$-string worldsheet theory and the corresponding $T$-dual modes in the $D3/D7$-$O7$ Type IIB orientifold theory.}}

We will explicitly construct and verify all of these zero-modes in the general supergravity background away from the orientifold limit in section \ref{Section3.5}.  Having deduced the existence of the 3-3 chiral fermionic zero-modes in the general case from index theory, let us first return to the anomaly puzzle and try to resolve it.

\subsection{Anomaly cancellation in the compact case}

We saw in section \ref{lackof} that the zero-mode anomaly from the 3-7 strings, $q_L$, cancels the anomaly inflow onto the $D3/D7$ intersection.  Now let us consider the anomalies of the 3-3 zero-modes.  Recall that the classically preserved symmetries of the system (and hence the ones that are potentially anomalous) are
\begin{eqnarray} SO(1,1) \times SO(6) \times O(N_c) \times G_f \ , \end{eqnarray}
where $SO(1,1)$ is the structure group of the tangent bundle of the intersection string, $T_{SO(1,1)}$, $SO(6)$ is the structure group of the normal bundle, $N_{SO(6)}$, of the string in $\mathbbm{R}^{1,7}$, $O(N_c)$ is the globally preserved gauge group of the $D3$-branes after taking into account the ``Alice string'' projection, and $G_f$ is the gauge group of the 7-branes.  The 3-3 zero-modes transform under these symmetries as indicated in the previous section.  Hence the associated anomaly polynomials are
\begin{eqnarray} I_{4}^{\xi_L} & = & \frac{1}{2} ch S(N_{SO(6)}) \wedge ch_{\frac{\mathbf{N}_{c} ( \mathbf{N}_c - \mathbf{1})}{\mathbf{2}}} (F_c) \wedge \hat{A}(T_{SO(1,1)}) |_{4} \ , \\
I_{4}^{\xi_R} & = & -\frac{1}{2} ch S(N_{SO(6)}) \wedge ch_{\frac{\mathbf{N}_{c} ( \mathbf{N}_c + \mathbf{1})}{\mathbf{2}}} (F_c) \wedge \hat{A}(T_{SO(1,1)}) |_{4} \ , \end{eqnarray}
where $S(N)$ denotes the spinor bundle of $N$.  We can evaluate these explicitly using
\begin{eqnarray} ch_{\frac{\mathbf{N}_{c} ( \mathbf{N}_c - \mathbf{1})}{\mathbf{2}}} (F_c) & = & \frac{ N_c (N_c - 1)}{2} + (N_c - 2) c_2(F_c) + \cdots \ , \\
ch_{\frac{\mathbf{N}_{c} ( \mathbf{N}_c + \mathbf{1})}{\mathbf{2}}} (F_c) & = & \frac{ N_c (N_c + 1)}{2} + (N_c + 2) c_2(F_c) + \cdots \ , \\
ch S(N_{SO(6)}) & = &  8 + p_1(N_{SO(6)}) + \cdots \end{eqnarray}
to find
\begin{eqnarray} I_{4}^{\xi_L} & = & \frac{N_c (N_c -1)}{4} p_1(N) + 4 (N_c - 2) c_2(F_c) - \frac{N_c(N_c -1)}{12} p_1(T), \\
I_{4}^{\xi_R} & = & - \frac{N_c (N_c + 1)}{4} p_1(N) - 4 (N_c + 2) c_2(F_c) + \frac{N_c(N_c +1)}{12} p_1(T). \end{eqnarray}
Adding these together gives
\begin{eqnarray} I_{4}^{\xi_L} + I_{4}^{\xi_R} = - \frac{N_c}{2} p_1(N_{SO(6)}) + \frac{N_c}{6} p_1(T_{SO(1,1)}) - 16 c_2(F_c). \end{eqnarray}
Observe that this is precisely equal and opposite to four times the inflow onto the intersection of the $D3$-branes with an $O7$-plane \eqref{D3O7Inflow}:
\begin{eqnarray} I_{4}^{\xi_L} + I_{4}^{\xi_R} + 4  I_{4}^{inf.D3/O7} & = & 0 \ . \end{eqnarray}
(We are ignoring the $SO(2)$ term in the inflow because it is not a symmetry of the configuration we are considering).  In the compact case, there are indeed four $O7$-planes.  Thus anomaly cancellation can be achieved if the 3-3 zero-modes are symmetrically localized on each of the $O7$-planes, such that their net anomaly is divided equally in four ways.  Studying the question of localization requires some knowledge of the explicit zero-mode wavefunctions.  Let us now turn to their construction.

\subsection{Explicit zero-mode solutions}  \label{Section3.5}

Using the effective action presented in Section \ref{Section3.2}, one can explicitly solve for both the bosonic and fermionic zero-modes in the general case, away from the orientifold limit.  One does indeed find unique solutions with the appropriate quantum numbers, corresponding to the content in Table \ref{Table2}.  The analysis is detailed and the arguments are subtle at some points, so we feel it is best to leave the explicit computations to the appendix.  Here we will emphasize one key point that is crucial to the analysis, and we present the results in Table \ref{Table3}.

The general strategy for finding massless modes localized along the $D3/O7$ intersection is straightforward.  One makes the ansatz $\Phi_{D3} = \phi(x^\mu) y(z,\bar{z})$, where $\Phi_{D3}$ is some $D3$-brane field, and solves for $y$ such that the $3+1$-dimensional equation of motion for $\Phi_{D3}$ is reduced to a $1+1$-dimensional wave equation for $\phi(x^\mu)$.  One easily obtains the general solution for $y(z,\bar{z})$ in all cases.  The powerful tool that allows us to find a \emph{unique} solution is the requirement of $SL(2,\mathbbm{Z})$ covariance.  Generically, $y$ solves an equation $\mathcal{L} y = 0$ and the operator $\mathcal{L}$ depends on the axidilaton $\tau$.  In the supergravity background $\tau$ undergoes various $SL(2,\mathbbm{Z})$ monodromies around closed loops, and thus so does $\mathcal{L}$, $\mathcal{L} \rightarrow \mathcal{L}'$.  However, since the background was constructed by gauging $SL(2,\mathbbm{Z})$, it follows that $y$ must also transform in such a way that the equation of motion is invariant: $\mathcal{L} y = 0 \iff \mathcal{L}' y ' = 0$.  Requiring that $y$ transforms appropriately allows us to fix a unique solution.  Note that there is also the requirement that $\phi(x^\mu)$ be normalizable to represent a zero-mode.  It is a nontrivial check that the $y$ we fix by $SL(2,\mathbbm{Z})$ covariance lead to an action that has finite integral over the transverse $\mathbbm{C}\mathbbm{P}^1$.

\TABLE{ \label{Table3}
\begin{tabular}{r|l}
\hline
D1 content & zero-mode  \\
\hline
gauge field $a_\mu$ & $\begin{array}{c c c} A_+ & = & c_+ a_+(x^\mu) \\ A_- & = & \frac{c_-}{\tau_2} a_-(x^\mu) \end{array} $ \\
\hline
scalars $X^I$ & $\begin{array}{c c c} A_z & = & c_z g(z) X^z(x^\mu) \\ A_{\bar{z}} & = & \bar{c_z} \bar{g}(\bar{z}) X^{\bar{z}}(x^\mu) \\ M^{ij} & = & c_\alpha \rho^{\alpha ij} X^{\alpha}(x^\mu) \end{array} $ \\
\hline
fermions $\begin{array}{c} \lambda_L = \xi_{L}^i \oplus \xi_{Li}^\ast \\ \theta_R = \xi_{R}^i \oplus \xi_{Ri}^\ast \end{array}$ & $ \begin{array}{c c c} f_+ & = & \frac{c_\lambda}{\sqrt{\tau_2}} \displaystyle\biggl( \frac{g(z)}{\bar{g}(\bar{z})} \displaystyle\biggr)^{1/4} \\ f_- & = & c_{\theta} \displaystyle\biggl( \frac{ \bar{g}(\bar{z})}{g(z)} \displaystyle\biggr)^{1/4} \end{array}$
\end{tabular}
\caption{The 3-3 string zero-modes.  $g(z)$ is the holomorphic function appearing in the metric, $g(z) = \eta^2(\bar{\tau}(z))/(\prod_{n=1}^{24} (z-z_{i\infty}^{(n)})^{1/12})$, and the $c$ are normalization constants to be determined.  The $\rho^{\alpha i j}$ that relate $M^{ij}$ and $X^\alpha$ are the $SU(4)$ Clebsch-Gordan coefficients.}}

Now that we have the explicit 3-3 fermion zero-modes in hand, let us return to the question of anomaly cancellation one more time.  Globally, there is no question; the chiral zero-modes found above have the right Lorentz and gauge quantum numbers to cancel the anomaly inflow associated with the intersection of the $N_c$ $D3$-branes with the 4 $O7$-planes.  The issue is whether the anomaly cancellation takes place locally.  Let us first suppose that we are away from the orientifold limit.  The zero-mode solutions are not really localized at the intersection of the $D3$-branes with an $O7$-plane, as there is no $O7$-plane.  On the other hand, the solutions are localized symmetrically around the four sets of $4D7+O7$'s.  One can really say no more of the $D3/O7$ inflow calculation either.  The most we can say with what has been presented so far is that the calculations--zero-modes and anomaly inflow--are at least consistent with local anomaly cancellation.

We claim that the solutions presented above correctly give the zero-mode wavefunctions, even in the region of strong coupling.  This is because they were uniquely fixed by considerations in the perturbative region, where we know our analysis is valid.  On the other hand, we know that the anomalous couplings on the $D7$-brane and $O7$-plane that we used, \eqref{D7anomalouscouplings}, \eqref{O7anomalouscouplings}, do receive corrections in the local string coupling $\tau$.  The exact couplings are known, to some extent, thanks to the Heterotic/Type I duality.  See for instance \cite{Lerche} and references therein.  It would be interesting to see if completely local anomaly cancellation could be verified using these results, but we will not pursue this here.

Let us make a few comments about the orientifold limit.  The axidilaton becomes constant, so the connection $Q_m$ is trivial.  There is, however, still an $S^2$ monodromy around each of the $O7$-planes, and the metric is nontrivial.  It has the same form, where we replace
\begin{eqnarray} \prod_{n=1}^{24} (z - z_{i\infty}^{(n)})^{1/24} \rightarrow \prod_{k=1}^4 (z - z_{O7}^{(k)})^{1/4} \ . \end{eqnarray}
In fact, the solutions we derived above for the zero-modes are still valid in this limit.  They solve the equations of motion, have the correct monodromies, and are normalizable.  The factors of $\tau_2$ and $\eta(\bar{\tau})$ are now simply constants.  Hence, in this limit, we clearly see that the zero-modes are symmetrically localized at the $z_{O7}^{(k)}$. This of course is in agreement with what one finds from a standard perturbative string calculation
of the zero-modes in this limit.

Finally, let us briefly return to the noncompact cases, $N_f = 6,12$.  We remarked at the end of the index computation that, on noncompact spaces, there is in general a boundary term contribution to the index.  The boundary is the surface at infinity, and this contribution can be nonzero if the gauge configuration ($Q_m$ in our case) does not fall off fast enough.  We expect that this will be the situation for us.  In the $N_f = 6,12$ cases, $Q_m$ has nontrivial monodromy\footnote{This is nicely shown in the analysis of \cite{Berg}.  Referring to the bottom figure on page 26 of that paper, each set of $4D7 + O7$'s has a point $z_i$ associated with it, about which there is $SL(2,\mathbbm{Z})$ monodromy $S^2$.  In the noncompact cases these $z_i$ are taken to infinity.  Then there will be a monodromy of $S^2$ or $S^4$ around $z = \infty$ in the $N_f = 6,12$ cases respectively.  $Q_m$ transforms under the double cover of $SL(2,\mathbbm{Z})$, (see Appendix \ref{AppendixB}), and so will have monodromy $Q_m \rightarrow i Q_m, -Q_m$ respectively.} around $z = \infty$ and so $\oint_{S_{\infty}^1} Q_\phi \neq 0$.  When properly accounted for, we expect that the boundary contribution will cancel the bulk result in both cases, so that the index vanishes.

This expectation  can be verified by explicit examination of the candidate zero-modes.
Proceeding as before, the general solutions for the transverse wavefunctions would be of the same form and we would impose the same conditions from $SL(2,\mathbbm{Z})$ covariance.  This would uniquely fix candidate solutions--they would be the same ones as in Table \ref{Table3}, but with the ``24'' in $g(z)$ replaced by a ``6'' or ``12.''  Then one quickly sees, following the analysis in the next section, that these solutions fail to be normalizable.  Thus one can show, by direct computation, that there are no zero-modes.

We have argued that the index can be consistent with this, but what about anomaly cancellation?  If there are no 3-3 string zero-modes in the noncompact cases, what does one make of the anomaly inflow onto the $D3/O7$ intersection in these cases?  There must be anomalous boundary WZ couplings for $Dp$-branes extended in noncompact spaces.  We conclude that these terms must be present here and give a cancelling contribution to the anomaly inflow, though we do not perform the explicit computation.  It is amusing to note that if one were to take the fractional result of the bulk index computation and conclude that there is a ``fourth of a zero-mode'' or ``half of a zero-mode'' in the $N_f = 6,12$ cases, then the corresponding anomalies, $\frac{1}{4} I^{\xi_{L,R}}$ or $\frac{1}{2} I^{\xi_{L,R}}$, would cancel the ``bulk'' inflow onto the intersection of the $D3$-branes with one or two $O7$-planes respectively.  This is an example where the bulk contribution to the index matches with the bulk contribution to the inflow, and thus the boundary contribution to each must match as well.

\section{Application: moduli dependence of the heterotic string coupling}

In this section we evaluate the 3-3 string effective action, given in \eqref{33bosonic} and \eqref{33fermionic}, on the zero-mode solutions found above, and integrate over the transverse $\mathbbm{C} \mathbbm{P}^1$ to obtain the $1+1$-dimensional theory.  When combined with the 3-7 string action, \eqref{37action}, the result should be the $D1$-string worldsheet theory \eqref{D1action}.  However, since we have the explicit zero-modes, we will be able to carry out this procedure in the general case, away from the orientifold limit.  Thus we will learn how the $D1$-string Yang-Mills coupling, and by Heterotic/Type I duality, the heterotic string coupling, depend on the moduli $z_{i\infty}$.

The strategy will be the following.  We will fix the normalization constants by canonically normalizing the $\mu$-direction kinetic terms in the 3-3 string effective action.  Then integrating the 3- and 4-point couplings over the $\mathbbm{C}\mathbbm{P}^1$ will give us the coupling $g_1(z_{i\infty})$.  It will be a very nontrivial check that all of these coefficients can be expressed in terms of a single $g_1$.

Let us begin with the gauge field action.  We momentarily employ indices $i,j$ to run over $z,\bar{z}$.  Simply by writing $A_m = (A_\mu, A_i)$, integrating the Chern-Simons-like term by parts, and using $\partial \tau = 0$, one can show
\begin{eqnarray} - \frac{1}{2\pi} \int d^4 x \sqrt{-g} \textrm{tr} \displaystyle\biggl( \frac{1}{4} e^{-\Phi} F_{mn} F^{mn} + \frac{1}{8} C_0 \epsilon^{mnpq} F_{mn} F_{pq} \displaystyle\biggr) =  \qquad \qquad \qquad \nonumber \\
\qquad \qquad =  -\frac{1}{2 \pi} \int d^4 x \tau_2 \sqrt{-g} \textrm{tr} \displaystyle\biggl( \frac{1}{4} F_{\mu\nu} F^{\mu\nu} + \frac{1}{2} \mathcal{D}_{\mu} A_i \mathcal{D}^\mu A^i + \frac{1}{4} F_{ij} F^{ij} + \nonumber \\
 \qquad \qquad + \frac{1}{2} ( \partial_i A_\mu \partial^i A^\mu + 2 \partial_i A_\mu [ A^i,A^\mu ] ) + \frac{1}{2} \epsilon^{\mu\nu} A_\mu \epsilon^{ij} (2 i Q_i) \mathcal{D}_j A_\nu \displaystyle\biggr). \end{eqnarray}
Evaluating this on the zero-mode solutions in Table \ref{Table3} completely kills the last line, including the 3-point couplings, and reduces $F_{ij} F^{ij} \rightarrow - [A_i,A_j] [A^i,A^j]$.  Hence one is left with
\begin{eqnarray} -\frac{1}{2 \pi} \int d^4 x \tau_2 \sqrt{-g} \textrm{tr} \displaystyle\biggl( \frac{1}{4} F_{\mu\nu} F^{\mu\nu} + \frac{1}{2} \mathcal{D}_{\mu} A_i \mathcal{D}^\mu A^i - \frac{1}{4} [A_i, A_j] [A^i, A^j] \displaystyle\biggr). \nonumber \\ \end{eqnarray}
Canonically normalizing the kinetic terms\footnote{Actually, canonically normalizing the kinetic terms for $a_+,a_-$ only fixes the product $c_+ c_- = (vol_{\mathbbm{C}\mathbbm{P}^1})^{-1}$.  We require that the coefficients of the 3-point couplings $a_+ [a_+, a_-]$ and $a_- [a_+, a_-]$ be the same in order to fix them individually.} leads to
\begin{eqnarray} c_- = \frac{1}{\sqrt{I_1}} \ , \qquad c_+ = \frac{\sqrt{I_1}}{vol_{\mathbbm{C}\mathbbm{P}^1}} \ , \qquad c_z = \frac{1}{\sqrt{vol_{\mathbbm{C}\mathbbm{P}^1}}} \ , \end{eqnarray}
where
\begin{eqnarray} vol_{\mathbbm{C}\mathbbm{P}^1} = \frac{-i}{2} \int_{\mathbbm{C}} dz d\bar{z} e^a \ , \qquad I_1 = \frac{-i}{2} \int_{\mathbbm{C}} dz d\bar{z} \frac{e^a}{\tau_2} \ . \end{eqnarray}
We then nontrivially find that all the 3-point couplings have coefficient $g_{1}$ and all the 4-point couplings have coefficient $g_{1}^2$, where
\begin{eqnarray} g_{1} =  \frac{\sqrt{I_1}}{vol_{\mathbbm{C}\mathbbm{P}^1}} \ . \label{coupling1} \end{eqnarray}
Thus, the gauge field action reduces to
\begin{eqnarray} S_I & = & -\frac{1}{2\pi} \int d^2 x \textrm{tr} \displaystyle\biggl( \frac{1}{4} F_{(2)\mu\nu} F_{(2)}^{\mu\nu} + \frac{1}{2} (\mathcal{D}_{(2)\mu} X^{i})^2 - \frac{g_{1}^2}{4} [X^i, X^j]^2 \displaystyle\biggr),  \end{eqnarray}
with $F_{(2)\mu\nu} = \partial_\mu a_\nu - \partial_\nu a_\mu + i g_{1} [a_\mu, a_\nu ]$ and $\mathcal{D}_{(2)\mu} = \partial_\mu + i g_{1} [a_\mu, \quad ]$.

The rest of the bosonic Lagrangian is $\frac{1}{2}(\mathcal{D}_m M_{ij})^2 - \frac{1}{4\tau_2} [M_{ij},M_{kl}]^2$.  We evaluate this on the zero-mode solution and find the normalization constants
\begin{eqnarray} c_{\alpha} = \frac{1}{\sqrt{ vol_{\mathbbm{C} \mathbbm{P}^1}} } \ . \end{eqnarray}
Then the 3- and 4-point couplings take the expected form with $g_1$ given in \eqref{coupling1}.  Thus the action for the scalars reduces to
\begin{eqnarray} S_{II} & = & - \frac{1}{2\pi} \int d^2 x \textrm{tr} \displaystyle\biggl( \frac{1}{2} (\mathcal{D}_{(2)\mu} X^\alpha)^2 - \frac{g_{1}^2}{2} [X^i, X^\alpha ]^2 - \frac{g_{1}^2}{4} [ X^{\alpha}, X^{\beta} ]^2 \displaystyle\biggr). \end{eqnarray}
Adding $S_I$ and $S_{II}$ gives the bosonic part of the 3-3 zero-mode action
\begin{eqnarray} S_{\textrm{3-3}z.m.}^{bos.} & = & - \frac{1}{2\pi} \int d^2 x \textrm{tr} \displaystyle\biggl( \frac{1}{4} F_{(2)\mu\nu} F_{(2)}^{\mu\nu} + \frac{1}{2} (\mathcal{D}_{(2)\mu} X^I)^2 - \frac{g_{1}^2}{4} [X^I,X^J]^2 \displaystyle\biggr).  \label{33zmbosonic} \end{eqnarray}

Now we move to the 3-3 effective fermionic action.  First consider the $\bar{\psi} \gamma^m (\mathcal{D}_m + \frac{i}{2}Q_m) L\psi$ term evaluated on the zero-mode solution.  We find
\begin{eqnarray} \frac{i}{2\pi} \int d^4 x \tau_2 \sqrt{-g}  \textrm{tr} \displaystyle\biggl( \bar{\psi}_i \gamma^m ( \mathcal{D}_m + \frac{i}{2} Q_m ) L \psi^i \displaystyle\biggr) =  \qquad \qquad \qquad \qquad \nonumber \\
\qquad \qquad  =  - \frac{i}{2\pi} \int d^4 x \tau_2 \sqrt{-g} \textrm{tr} \displaystyle\biggl( |f_+|^2 \xi_{Li}^\dag \mathcal{D}_- \xi_{L}^i + |f_-|^2 \xi_{Ri}^\dag \mathcal{D}_+ \xi_{R}^i + \nonumber \\
+  i f_{+}^\ast f_- e^{-a/2} \xi_{Li}^\dag [A_{z}, \xi_{R}^i] + i f_{-}^\ast f_+ e^{-a/2} \xi_{Ri}^\dag [ A_{\bar{z}}, \xi_{L}^i ]  \displaystyle\biggr), \end{eqnarray}
where, recall,
\begin{eqnarray} f_+ = \frac{c_\lambda}{\sqrt{\tau_2}} \displaystyle\biggl( \frac{g(z)}{\bar{g}(\bar{z}) } \displaystyle\biggr)^{1/4} \ , \qquad f_- = c_\theta \displaystyle\biggl( \frac{ \bar{g}(\bar{z})}{ g(z)} \displaystyle\biggr)^{1/4} \ . \end{eqnarray}
Canonically normalizing the kinetic terms we find
\begin{eqnarray} c_\lambda = \frac{1}{\sqrt{ vol_{\mathbbm{C}\mathbbm{P}^1} }} \ , \qquad c_\theta = \frac{1}{\sqrt{I_2}} \ , \end{eqnarray}
where
\begin{eqnarray} I_2 = - \frac{i}{2} \int_{\mathbbm{C}} dz d\bar{z} \tau_2 e^{a} \ . \end{eqnarray}
Then we find that the coefficients of the $\xi_{Li}^\dag a_- \xi_{L}^i$ and $\xi_{Ri}^\dag a_+ \xi_{R}^i$ terms are both $g_{1}$.  \emph{However}, the coefficients of the Yukawa terms, $\xi_{Li}^\dag [A_{z}, \xi_{R}^i]$ and its conjugate, are both
\begin{eqnarray} g_{1}' = \frac{1}{\sqrt{I_2}} \ . \end{eqnarray}

We will address this difference shortly, but let us first finish the dimensional reduction.
The last terms are the Yukawa couplings of the fermions to the scalars $M^{ij}$.  Plugging in the zero-mode solutions and integrating over the transverse space leads to
\begin{eqnarray} \frac{i}{2\pi} \int d^4 x \sqrt{-g} \sqrt{\tau_2} \textrm{tr} \displaystyle\biggl( (L\psi)^i [ (L\psi)^j, M_{ij}] + (\bar{\psi} R)_i [ (\bar{\psi} R)_j, M^{ij}] \displaystyle\biggr)  = \qquad  \nonumber \\
\qquad = - \frac{i}{2\pi} \int d^2x \textrm{tr} \displaystyle\biggl( g_{1}' \rho^{\alpha}_{ij} \xi_{L}^i [ \xi_{R}^j, X^{\alpha} ] + g_{1}' \rho^{\alpha ij} \xi_{Li}^\dag [ \xi_{Rj}^\dag, X^\alpha ] \displaystyle\biggr). \end{eqnarray}
These terms also have the coupling constant $g_{1}'$ instead of $g_1$.  Hence, after some slight rearranging, we obtain
\begin{eqnarray} S_{\textrm{3-3}z.m.}^{ferm.} & = & - \frac{i}{2\pi} \int d^2 x \textrm{tr} \displaystyle\biggl( \xi_{Li}^\dag \mathcal{D}_{(2)-} \xi_{L}^i + \xi_{Ri}^\dag \mathcal{D}_{(2)+} \xi_{R}^i + i g_{1}' \xi_{Li}^\dag [\xi_{R}^i, X^z] \nonumber \\
& & + i g_{1}' \xi_{L}^i [ \xi_{Ri}^\dag, X^{\bar{z}} ] + g_{1}' \rho^{\alpha}_{ij} \xi_{L}^i [\xi_{R}^j , X^{\alpha} ] + g_{1}' \rho^{\alpha ij} \xi_{Li}^\dag  [\xi_{Rj}^\dag , X^\alpha ] \displaystyle\biggr) \nonumber \\ \label{D3O7fermion} \end{eqnarray}
for the 3-3 fermionic zero-mode action.

In order to put this action in an $SO(8)$ invariant form, we take a brief detour through some gamma matrix definitions.  Consider the embedding of SO(6) gamma matrices $\gamma^\alpha$ into $SO(8)$ gamma matrices $\Gamma^I$ according to
\begin{eqnarray} \Gamma^1 = \sigma^2 \otimes \mathbbm{1}_8 \ , \qquad \Gamma^2 = \sigma^1 \otimes \bar{\gamma} \ , \qquad \Gamma^{2+\alpha} = \sigma^1 \otimes \gamma^\alpha \ , \end{eqnarray}
where $\bar{\gamma} = -i \gamma^1 \cdots \gamma^6$.  It is straightforward to show that $\bar{\Gamma} = \prod \Gamma^I = \sigma^3 \otimes \mathbbm{1}_8$, so that the $SO(8)$ Dirac spinor decomposes as $\Psi = (\mathbf{8}_s, \mathbf{8}_c )^T$.  Then we define $\sigma^I$ by
\begin{eqnarray} \Gamma^I = \left( \begin{array}{c c} 0 & \sigma^I \\ (\sigma^I)^\dag & 0 \end{array} \right). \end{eqnarray}
It is clear that these may be viewed as Clebsch-Gordan coefficients for $\mathbf{8}_s \times \mathbf{8}_c \rightarrow \mathbf{8}_v$.  Now, the $\gamma^\alpha$ in turn may be represented using the $SU(4)$ Clebsch-Gordan coefficients,
\begin{eqnarray} \gamma^\alpha = \left( \begin{array}{c c} 0 & \rho^{\alpha ij} \\ \rho^{\alpha}_{ij} & 0 \end{array}\right), \end{eqnarray}
where, recalling the reality constraint, $\rho^{\alpha}_{ij} \equiv (\rho^{\alpha ij})^\dag = \frac{1}{2} \epsilon_{ijkl} \rho^{\alpha kl}$.  The $\rho^\alpha$ also satisfy a normalization condition $\rho^{\alpha ij} \rho^{\beta}_{jk} + \rho^{\beta ij} \rho^{\alpha}_{jk} = 2 \delta^{i}_{\phantom{i}k} \delta^{\alpha\beta}$; this ensures that the $\gamma^\alpha$ satisfy the Clifford algebra.  A specific choice for the $\rho^\alpha$ exists such that $\bar{\gamma} = \sigma^3 \otimes \mathbbm{1}_4$, and the $SO(6)$ charge conjugation matrix $C$, satisfying $C^{-1} \gamma^\alpha C = - \gamma^{\alpha T}$, $C^T = C^\ast = C$, is given by $C = \sigma^1 \otimes \mathbbm{1}_4$.  This is the charge conjugation matrix that appears in \eqref{D1action}.  The $SO(8)$ charge conjugation matrix may be taken as $\mathcal{C} = \mathbbm{1}_2 \otimes C$.

{}From these results it follows that
\begin{eqnarray} \lambda_L = \frac{1}{\sqrt{2}} \left(\begin{array}{c} \xi_{L}^i \\ \xi_{Li}^\ast \end{array}\right), \qquad \theta_{R} = \frac{1}{\sqrt{2}} \left(\begin{array}{c} \xi_{R}^i \\ \xi_{Ri}^\ast \end{array}\right) \end{eqnarray}
are Weyl spinors in the $\mathbf{8}_s, \mathbf{8}_c$ respectively and both self-conjugate, $\lambda_{L}^\ast = C \lambda_L, \theta_{R}^\ast = C \theta_R$.  It is also straightforward to show that \eqref{D3O7fermion} may be written as
\begin{eqnarray}  S_{\textrm{3-3}z.m.}^{ferm.} & = & - \frac{i}{2\pi} \int d^2 x \textrm{tr} \displaystyle\biggl( \lambda_{L}^T C \mathcal{D}_{(2)-} \lambda_L + \theta_{R}^T C \mathcal{D}_{(2)+} \theta_R + 2 g_{1}' \lambda_{L}^T C \sigma^I [\theta_R, X^I] \displaystyle\biggr). \label{33zmfermionic} \nonumber  \\ \end{eqnarray}

Now observe that after summing the bosonic and fermionic zero-mode actions, and the 3-7 action, \eqref{33zmbosonic},\eqref{33zmfermionic},\eqref{37action}, and rescaling the gauge multiplet $(\lambda_L, a_\mu) \rightarrow \frac{1}{g_{1}} (\lambda_L, a_\mu)$, we obtain the $D1$-string worldsheet action \eqref{D1action}, if and only if the two couplings $g_1,g_{1}'$ are the same:
\begin{eqnarray} S_{\textrm{3-3}z.m.}^{bos.} + S_{\textrm{3-3}z.m.}^{ferm.} + S_{\textrm{3-7}} = S_{D1} \qquad \iff \qquad g_1 = g_{1}' \ . \end{eqnarray}
It is well known that the $(0,8)$ theory \eqref{D1action} is completely specified by the gauge coupling.  If the Yukawa term $\lambda \theta X$ had a different coefficient, supersymmetry would be broken.  On the other hand, we know that the general background we have been studying preserves these 8 supercharges.  Hence it must in fact be the case that $g_1 = g_{1}'$.  Let us rephrase this equality slightly.  Let us define the average of a quantity over the $\mathbbm{C}\mathbbm{P}^1$ as
\begin{eqnarray} \langle x \rangle = \frac{ \int_{\mathbbm{C}} d^2z e^a x }{\int_{\mathbbm{C}} d^2z e^a } \ . \end{eqnarray}
Then we have
\begin{eqnarray} g_{1}^2 = \frac{\langle \tau_{2}^{-1} \rangle }{vol_{\mathbbm{C}\mathbbm{P}^1}} \ , \qquad g_{1}'^2 = \frac{\langle \tau_2 \rangle^{-1} }{vol_{\mathbbm{C}\mathbbm{P}^1} } \label{g1def} \end{eqnarray}
and the claim is that $\langle \tau_{2}^{-1} \rangle = \langle \tau_{2} \rangle^{-1}$.  In the general case (general elliptically fibered K3), $\tau$ is a very complicated function on the base $\mathbbm{C}\mathbbm{P}^1$.  We do not know how to do either of these integrals or why this should be true, yet supersymmetry leads us to conjecture that it is so!  Note, however, that in the orientifold limit, where $\tau_2$ becomes constant, it is trivially true.

Now, from the $D1$-string point of view, $2\pi g_{1}^2$ is, by definition, the square of the $D1$-brane Yang-Mills coupling, which is proportional to the Type I string coupling $g_{s,I}$.  We have $g_{1}^2 = g_{s,I}/(4\pi^2\alpha')$.  On the other hand, equation \eqref{g1def} is giving $g_{1}^2$ as a function of of the 7-brane, or F-theory on K3 moduli $(z_{i\infty},g_{s,I'})$, (where $g_{s,I'} = \lim_{z\rightarrow \infty} \tau_{2}^{-1}$).  Thus, our result gives part of the map between moduli spaces of Type I on $T^2$ and F-theory on $K3$.  The duality of these two theories is argued by going to a special point in the moduli space, corresponding to the orientifold limit of the 7-brane system, where they are $T$-dual.  It is easy to check that our result reduces to the well known map there.  In the orientifold limit, $\tau_2 = 1/g_{s,I'} = const$ and $\mathbbm{C}\mathbbm{P}^1$ becomes a torus.  Thus our result follows easily from the usual relations \cite{PolchinskiWitten}
\begin{eqnarray} \frac{R_{I',1} R_{I',2}}{g_{s,I'}^2} =  \frac{R_{I,1} R_{I,2}}{g_{s,I}^2} \ , \qquad R_{I,i} = \frac{\alpha'}{R_{I',i}} \ . \label{Tmap} \end{eqnarray}
In the orientifold limit, it is also possible to explicitly evaluate $vol_{\mathbbm{C}\mathbbm{P}^1}$, and hence $g_{1}^2$, as a function of the four $z_{O7}^{(k)}$.  In fact, one can use the $SL(2,\mathbbm{C})$ coordinate freedom to conveniently fix three of the $O7$-plane locations.  If one takes $(z_{O7}^{(1)},\ldots,z_{O7}^{(4)}) = (0,z^\ast,1,\infty)$, with $z^\ast$ arbitrary, then the result \cite{LercheStieberger} is
\begin{eqnarray} vol_{\mathbbm{C}\mathbbm{P}^1} & = &   Im \displaystyle\biggl( \frac{1}{\sqrt{z^\ast}} \frac{ {}_2 F_1(\frac{1}{2},\frac{1}{2},1; 1/z^\ast) }{ {}_2 F_1 (\frac{1}{2}, \frac{1}{2}, 1; z^\ast) } \displaystyle\biggr). \end{eqnarray}

There is also much evidence for the conjectured strong/weak duality between heterotic and Type I, where the coupling constants are related by $g_{s,h} = 1/g_{s,I}$.  Indeed, it is convincingly argued in \cite{Lowe,Rey} that the $D1$-string theory \eqref{D1action} flows to a strong-coupling superconformal fixed point, corresponding to fundamental heterotic strings on a certain orbifold space.  Thus we learn how the heterotic string coupling depends on the F-theory moduli,
\begin{eqnarray} g_{s,h}(z_{i\infty},g_{s,I'}) = \frac{1}{4\pi^2 \alpha' g_{1}^2} = \frac{vol_{\mathbbm{C} \mathbbm{P}^1}}{4\pi^2 \alpha' \langle \tau_{2}^{-1} \rangle } \ . \end{eqnarray}

This is part of the map between the moduli spaces of F-theory compactified on an elliptically fibered K3 and heterotic compactified on $T^2$.  It has been proven in \cite{ClingherMorgan} that the classical moduli spaces of these two theories are isomorphic.  They showed that both have the structure of a certain holomorphic $\mathbbm{C}^\ast$ fibration, which leads to a natural isomorphism between the two spaces.  The classical moduli spaces should well approximate the full (quantum) moduli spaces in regions where the quantum effects are negligible.  These regions correspond to large volume on the heterotic side and being near the orientifold limit on the F-theory side.  This is precisely the regime where our analysis is valid.  We required $g_{s,I'}$ small in order to have a field theory description of the $D3$-branes.  Combining \eqref{Tmap} with the Type I/heterotic relation $R_{I,i} = R_{h,i} g_{s,I}^{1/2}$ implies that
\begin{eqnarray} g_{s,I'} = \frac{\alpha'}{R_{h,1} R_{h,2}} \ . \end{eqnarray}
Thus $g_{s,I'}$ small does correspond to the volume of the heterotic torus being large.


Our map should agree with the one presented in \cite{ClingherMorgan}.  However, their parametrization of the F-theory moduli space is in terms of periods of the holomorphic two-form on the K3.  It would be interesting to explore the relation between these two parametrizations.

Finally, the map of \cite{ClingherMorgan} is more complete, as it specifies how the rest of the heterotic string moduli, namely the Wilson lines, depend on the K3 data.  Indeed, we have neglected a term in our effective action.  Recall that the heterotic sigma model action in a general background with nontrivial gauge field has a coupling of the current algebra fermions to the pullback of the gauge field to the worldsheet, roughly $\bar{\lambda} \partial X A \lambda$.  There should be a corresponding term in the $D1$-string action, which would involve a coupling of the $\chi$ to the $X^I$ and the background $D7$-brane gauge field.  This term would be present if the gauge field has nontrivial Wilson lines, and so it should be present for us since, roughly, the Wilson lines correspond to the location of the $D7$-branes.  One should be able to derive this term in the $D3/D7$-$O7$ picture by computing a disk amplitude with one 3-3, one 7-7, and two 3-7 boundary vertex operators.  It would be interesting to do this and see if one can derive the full map of F-theory and heterotic moduli spaces.

\section{Discussion}

In this paper we have have studied in detail the structure of $D3$-branes intersecting $D7$-branes
and $O7$-planes in $1+1$-dimensions. We used anomaly arguments, index theory, and an explicit
construction of the zero-modes to show that the $D3$-brane in this background has zero-modes
which are localized near the $O7$-planes and, in the compact case, fill out the multiplets which
gives rise to the zero-modes of the heterotic matrix string.  Our results explicitly show that the $D3$-brane zero-modes are not present in the noncompact cases.  In the compact case, by evaluating the $D3$-brane effective action on the zero-modes and integrating over the space transverse to the $D7$-branes and $O7$-planes, we obtained the $1+1$-dimensional theory of $N_c$ coincident $D1$-strings in Type I.  By carrying out this procedure away from the orientifold limit, we learned how the Type I string coupling, and by Type I/heterotic duality, the heterotic string coupling depend on the 7-brane moduli, or equivalently the moduli of F-theory compactified on K3.

In work to appear we plan to study this system at strong 't Hooft coupling where it is described
by supergravity on $AdS_3$, in particular by F-theory on $AdS_3 \times S^5 \times K3$.
This provides an explicit $AdS_3$ dual of the world-sheet CFT of $N_c$ heterotic strings,
albeit at strong string coupling, and should be a useful tool in trying to resolve some of the
puzzles raised recently concerning the structure of this duality \cite{Lapan,Dabholkar,Johnson,Kraus}.
AdS/CFT duals of matrix string theory have been discussed previously in \cite{Morales}.

\acknowledgments{The work of JH and AR  was supported in part by NSF Grant No. PHY-00506630 and NSF Grant 0529954. Any opinions, findings, and conclusions or recommendations expressed in this material are those of the authors and do not necessarily reflect the views of the National Science Foundation. AR acknowledges support from GAANN grant P200A060226 and the hospitality of the University of Cincinnati during the completion of this work. JH thanks the Aspen Center for Physics for providing a stimulating environment during parts of this research.}

\appendix
\section{$D3$-brane fermionic effective action} \label{AppendixA}

We begin with the action
\begin{eqnarray} S_{D3}^{ferm.} & = & \frac{i T_{D3}}{2} \int d^4 \xi \sqrt{-g} \textrm{tr} \displaystyle\biggl( \bar{y} (1 - \tilde{\Gamma}_{D3}) (e^{-\Phi/4} \hat{\Gamma}^m \breve{D}_m - \breve{\Delta} ) y \displaystyle\biggr). \end{eqnarray}
Let us review what these various quantities are, following \cite{MMS}.  We denote by $m,n,\ldots$ and $a,b,\ldots$ worldvolume/tangent space indices along the brane, and $M,N,\ldots$, $A,B,\ldots$ denote spacetime/tangent space indices in the bulk.  We have
\begin{eqnarray} y = \left(\begin{array}{c} y_1 \\ y_2 \end{array}\right), \end{eqnarray}
where $y_1,y_2$ are each 32-component $d=10$ spinors, satisfying Majorana and Weyl constraints, both of the same chirality.  Also,
\begin{eqnarray} \hat{\Gamma}^A = \left( \begin{array}{c c} \Gamma^A & 0 \\ 0 & \Gamma^A \end{array}\right) = I_2 \otimes \Gamma^A \ , \qquad \hat{\bar{\Gamma}} = \left(\begin{array}{c c} \bar{\Gamma} & 0 \\ 0 & - \bar{\Gamma} \end{array}\right) = \sigma^3 \otimes \bar{\Gamma} \ , \end{eqnarray}
where $\Gamma^A$ are the $d=10$ gamma matrices and $\bar{\Gamma}$ the generalized ``$\gamma^5$.''  $\tilde{\Gamma}_{D3}$ is given by
\begin{eqnarray} \tilde{\Gamma}_{D3} = -i \sigma^2 \otimes \frac{1}{4! \sqrt{-g}} \varepsilon^{m_1 \cdots m_4} \Gamma_{m_1 \cdots m_4} \ , \end{eqnarray}
where $\varepsilon^{0123} = 1$ and $\Gamma_{m_1 \cdots m_4} = \Gamma_{[m_1} \cdots \Gamma_{m_4]}$.  $\frac{1}{2}(1 - \tilde{\Gamma}_{D3})$ is a kappa symmetry projection operator that will remove half of the degrees of freedom.  Finally, $\breve{D}_m$ and $\breve{\Delta}$ are given by
\begin{eqnarray} \breve{D}_m = \mathbbm{1}_2 \otimes \hat{D}^{(0)}_m + \sigma^1 \otimes \hat{W}_m \ , \qquad \breve{\Delta} = \mathbbm{1}_2 \otimes \hat{\Delta}^{(1)} + \sigma^1 \otimes \hat{\Delta}^{(2)} \ , \end{eqnarray}
where
\begin{eqnarray} \hat{D}_{(1,2)M}^{(0)} & = & \partial_M + \frac{1}{4} \displaystyle\biggl( \omega_{AB,M} + \frac{1}{4} \tau_{AB,M} \displaystyle\biggr) \Gamma^{AB} \equiv \tilde{D}_M \ , \\
\hat{W}_{(1,2)M} & = & \frac{1}{8} \displaystyle\biggl( \mp e^{3 \Phi/4} G^{(1)}_{A} \Gamma^A \displaystyle\biggr) e^{ \Phi/4} \Gamma_M \ , \\
\hat{\Delta}_{(1,2)}^{(1)} & = & \frac{1}{2} e^{- \Phi/4} \Gamma^M \partial_M \Phi \ , \\
\hat{\Delta}_{(1,2)}^{(2)} & = & \pm \frac{1}{2} e^{3 \Phi/4} G^{(1)}_{A} \Gamma^A \ . \end{eqnarray}
The subscript $(1,2)$ is correlated with the sign.  If the operator acts on $y_1$ the top sign is chosen, if it acts on $y_2$ the bottom sign is chosen.

One important step has already been taken relative to the formulae presented in \cite{MMS}.  Their results are given in string frame and we have converted to Einstein frame using $g_{MN}^{(s)} = e^{\Phi/2} g_{MN}^{(e)}$.  In terms of the vielbeins and inverse vielbeins, $e^{(s)A}_{\phantom{(s)A}M} = e^{\Phi/4} e^{(e)A}_{\phantom{(s)A}M}$ and $E_{A}^{(s)M} = e^{-\Phi/4} E_{A}^{(e)M}$.  Thus, for instance, one has $\Gamma^{(s)M} = e^{-\Phi/4} \Gamma^{(e)M}$ and $G_{A}^{(1)(s)} = e^{-\Phi/4} G_{A}^{(1)(e)}$.  The projector $\tilde{\Gamma}_{D3}$ is unchanged; the transformation of the vielbeins in $\Gamma_{m_1 \cdots m_4}$ cancels the factor coming from $\sqrt{-g}$ in the denominator.  Finally, it can be shown that the spin connection gets modified: $\omega_{AB,M}^{(s)} = \omega_{AB,M}^{(e)} + \frac{1}{4} \tau_{AB,M}^{(e)}$, where $\tau_{AB,M}$ is defined through
\begin{eqnarray} \label{taudef1}  \tau_{ab,c} & = & \frac{1}{2} ( \chi_{a,bc} + \chi_{b,ca} - \chi_{c,ab} ) \ , \end{eqnarray}
with $\chi_{a,bc} = -\chi_{a,cb}$ given by
\begin{eqnarray} \label{taudef2}  d\Phi \wedge e_a & = & \frac{1}{2} \chi_{a,bc} e^b \wedge e^c \ . \end{eqnarray}

Now let us simplify this action.  We have
\begin{eqnarray} \frac{1}{4! \sqrt{-g}} \varepsilon^{m_1 \cdots m_4} \Gamma_{m_1 \cdots m_4} = \frac{e^{a_1}_{\phantom{a_1}m_1} \cdots e^{a_4}_{\phantom{a_4}m_4} }{4! \sqrt{-g}} \varepsilon^{m_1 \cdots m_4} \Gamma_{a_1 \cdots a_4} = \Gamma_{\underline{0} \underline{1} \underline{2} \underline{3}} = - i \bar{\Gamma}_{(4)} \ , \nonumber \\ \end{eqnarray}
where we've defined $\bar{\Gamma}_{(4)} = -i \Gamma^{\underline{0} \cdots \underline{3}}$ which anticommutes with all of the $\Gamma^a$ and squares to one.  Thus
\begin{eqnarray} 1 - \tilde{\Gamma}_{D3} = \left(\begin{array}{c c} 1 & - i \bar{\Gamma}_{(4)} \\ - \bar{\Gamma}_{(4)} & 1 \end{array}\right). \end{eqnarray}
Using the fact that the background supergravity fields only depend on directions tangent to the brane worldvolume, we eventually find
\begin{eqnarray} e^{- \Phi/4} \Gamma^m \breve{D}_m - \breve{\Delta} & = &  e^{-\Phi/4}  \left( \begin{array}{c c}  \Gamma^m \tilde{D}_m - \frac{1}{2} \Gamma^m \partial_m \Phi & \frac{1}{4} e^{\Phi} G^{(1)}_{a} \Gamma^a \\ - \frac{1}{4} e^{ \Phi} G^{(1)}_{a} \Gamma^a & \Gamma^m \tilde{D}_m - \frac{1}{2} \Gamma^m \partial_m \Phi \end{array} \right).  \end{eqnarray}
Then one can show
\begin{eqnarray} (1 - \tilde{\Gamma}_{D3}) ( e^{- \Phi/4} \Gamma^m \breve{D}_m - \breve{\Delta} ) & = & e^{-\Phi/4} \left(\begin{array}{c c} \Gamma^m \mathcal{\tilde{D}}_m & \Gamma^m \mathcal{\tilde{D}}_m (i \bar{\Gamma}_{(4)}) \\ \Gamma^m \mathcal{\tilde{D}}_m (- i \bar{\Gamma}_{(4)}) & \Gamma^m \mathcal{\tilde{D}}_m \end{array}\right) \nonumber \\
& \equiv & e^{-\Phi/4} \mathcal{M} \ ,  \end{eqnarray}
where
\begin{eqnarray} \mathcal{\tilde{D}}_m & = & \tilde{D}_m - \frac{1}{2} \partial_m \Phi - \frac{i}{4} e^{\Phi} G_{m}^{(1)} \bar{\Gamma}_{(4)} \ . \end{eqnarray}

Now consider the unitary change of variables $\tilde{y} = U y$ given by
\begin{eqnarray} \left( \begin{array}{c} \tilde{y}_1 \\ \tilde{y}_2 \end{array}\right) = \frac{1}{\sqrt{2}} \left( \begin{array}{c c} \bar{\Gamma}_{(4)} & -i \bar{\Gamma}_{(4)} \\ \bar{\Gamma}_{(4)} & i \bar{\Gamma}_{(4)} \end{array} \right) \left( \begin{array}{c} y_1 \\ y_2 \end{array}\right). \end{eqnarray}
Then observe that
\begin{eqnarray} U \mathcal{M} U^\dag = - \left( \begin{array}{c c} \Gamma^m \mathcal{\tilde{D}}_m (1 - \bar{\Gamma}_{(4)} ) & 0 \\ 0 & \Gamma^m \mathcal{\tilde{D}}_m ( 1 + \bar{\Gamma}_{(4)}) \end{array} \right) \equiv - \tilde{\mathcal{M}} \ . \end{eqnarray}
Therefore $\bar{y} \mathcal{M} y = \bar{\tilde{y}} \tilde{\mathcal{M}} \tilde{y}$.  Now note that, if $y_1,y_2$ are both left-handed, then $\tilde{y}_1 = \bar{\Gamma}_{(4)} ( y_1 + i y_2)$ and $\tilde{y}_2 = \bar{\Gamma}_{(4)} (y_1 - i y_2)$ are also both left-handed since $\bar{\Gamma}_{(4)}$ commutes with $\bar{\Gamma}$.  On the other hand, if the original $y_1,y_2$ are each separately Majorana, $y_1 = C \bar{y}_{1}^{T}$ and $y_2 = C \bar{y}_{2}^{T}$, we now have that $C \bar{\tilde{y}}_{1}^{T} = \pm \tilde{y}_2$ and $C \bar{\tilde{y}}_{2}^T = \pm \tilde{y}_1$, with the sign depending on whether $\bar{\Gamma}_{(4)}$ commutes or anticommutes with $C (\Gamma^0)^T$.

Now let us pick a convenient basis for the $\Gamma^A$ and dimensionally reduce to $d = 4$.  Let
\begin{eqnarray} \Gamma^{0,\ldots,3} = \mathbbm{1}_8 \otimes \gamma^{0,\ldots,3} \ , \qquad \Gamma^{4,\ldots,9} = \rho^{1,\ldots,6} \otimes \bar{\gamma} \ , \end{eqnarray}
where
\begin{eqnarray} \{ \gamma^a, \gamma^b \} = 2 \eta^{ab} \mathbbm{1}_4 \ , \qquad \{ \rho^{\alpha}, \rho^{\beta} \} = 2 \delta^{\alpha \beta} \mathbbm{1}_8 \end{eqnarray}
are $d=4$ Minkowski and $d=6$ Euclidian gamma matrices respectively.  There exists a nice basis\footnote{See \cite{Sohnius} for instance.} for the $\rho^{\alpha}$ such that the four-dimensional Weyl projector and charge conjugation matrix are related the their ten-dimensional counterparts by
\begin{eqnarray} L_{(10)} = \left(\begin{array}{c c} \mathbbm{1}_4 \otimes L_{(4)} & 0 \\ 0 & \mathbbm{1}_4 \otimes R_{(4)} \end{array}\right), \qquad C = \left(\begin{array}{c c} 0 & \mathbbm{1}_4 \otimes c \\ \mathbbm{1}_4 \otimes c & 0 \end{array} \right). \end{eqnarray}
Then, writing
\begin{eqnarray} \tilde{y}_1 = \left( \begin{array}{c} \psi_{(1)}^i \\ \chi_{(1) i} \end{array} \right), \qquad \tilde{y}_2 = \left( \begin{array}{c} \psi_{(2)}^i \\ \chi_{(2) i} \end{array} \right) \end{eqnarray}
with $i = 1,\ldots 4$ the $SU(4)_R$ index and each $\psi^i,\chi_i$ a $d=4$ four-component spinor, one sees that the ten-dimensional Weyl and Majorana conditions reduce to
\begin{eqnarray} \begin{array}{c c c} L \psi_{(1,2)} & = & \psi_{(1,2)} \\ R \chi_{(1,2)} & = & \chi_{(1,2)} \end{array} \qquad \textrm{and} \qquad \begin{array}{c c c} c \bar{\psi}_{(1)}^T & = & \chi_{(2)} \\ c \bar{\psi}_{(2)}^T  &= & \chi_{(1)} \end{array} \ . \label{spinorrelations} \end{eqnarray}
Recalling the form of $\tilde{\mathcal{M}}$, and noting that $\bar{\Gamma}_{(4)} = \mathbbm{1}_8 \otimes \bar{\gamma}$, one now sees that $\psi_{(1)}$ and $\chi_{(2)}$ are projected out.  This is the $\kappa$-projection working.  The action boils down to
\begin{eqnarray} S_{D3}^{ferm.} & = & \frac{i T_{D3}}{2} \int d^4 x \sqrt{-g} e^{-\Phi/4} \textrm{tr} \displaystyle\biggl( \bar{\chi}_{(1)}^i \gamma^m \mathcal{\tilde{D}}_{m}^- R \chi_{(1)i} + \bar{\psi}_{(2)i} \gamma^m \mathcal{\tilde{D}}_{m}^+ L \psi_{(2)}^i \displaystyle\biggr), \nonumber \\ \end{eqnarray}
where
\begin{eqnarray} \mathcal{\tilde{D}}_{m}^{\pm} & = & \tilde{D}_m - \frac{1}{2} \partial_m \Phi \mp \frac{i}{4} e^{\Phi} G_{m}^{(1)} \end{eqnarray}
comes from $\bar{\gamma} R = -R, \bar{\gamma} L = L$.

Now set the first term equal to minus its transpose and use the charge conjugation relations \eqref{spinorrelations}.  The minus comes from the fact that the fermions are Grassmann valued.  After carefully moving one of the resulting charge conjugation matrices to the other, one finds that the $\tilde{D}_m$ and $\frac{i}{4} e^{\Phi} G_{m}^{(1)}$ terms add,\footnote{The $\partial_m$ term gets a second minus from integration by parts.} while the $\frac{1}{2} \partial_m \Phi$ term cancels out.  Thus, defining $\psi^i \equiv \psi_{(2)}^i$, we obtain
\begin{eqnarray} S_{D3}^{ferm.} = T_{D3} \int d^4 x \sqrt{-g} e^{-\Phi/4} \textrm{tr} \displaystyle\biggl( \bar{\psi}_i \gamma^m ( i \tilde{D}_m + \frac{1}{4} e^{\Phi} G_{m}^{(1)} ) L \psi^i \displaystyle\biggr). \end{eqnarray}

{}From the definition of $\tau_{ab,m}$, \eqref{taudef1}, \eqref{taudef2}, it is straightforward to verify that
\begin{eqnarray} \frac{1}{16} \gamma^m \tau_{ab,m} \gamma^{ab} = \frac{3}{8} \gamma^m \partial_m \Phi \ . \end{eqnarray}
Thus by making the field redefinition $\psi \rightarrow e^{-3 \Phi/8} \psi$, we can cancel this factor in the Dirac operator and obtain an overall factor of $e^{-\Phi} = \tau_2$ out front.  Also, observe that
\begin{eqnarray} - \frac{i}{4} e^{\Phi} \partial_m C_0 = - \frac{i}{4} \frac{ \partial_m \tau_1}{\tau_2} = - \frac{i}{4} \frac{\partial_m (\tau + \bar{\tau})}{(-i)(\tau - \bar{\tau})} = \frac{i}{2} Q_m \ . \end{eqnarray}
Finally, the conventions of \cite{MMS} are such that $2\pi \alpha' = 1$, and therefore $T_{D3} = 1/2\pi$.  Putting all of this together, we obtain the desired result
\begin{eqnarray}  S_{\textrm{3-3}}^{ferm.} & = &  \frac{i}{2\pi} \int d^4 x \tau_2 \sqrt{-g}  \textrm{tr} \displaystyle\biggl( \bar{\psi}_i \gamma^m ( D_m + \frac{i}{2} Q_m ) L \psi^i \displaystyle\biggr) + O(\psi^2 A, \psi^2 M). \end{eqnarray}

\section{Explicit formulae for zero-modes} \label{AppendixB}

Having argued for the existence of 3-3 string zero-modes in the compact case using index theory, string duality, and anomaly cancelation, let us now explicitly construct them.  We will work in the general case with spacetime varying axidilaton, away from the orientifold limit.

\subsection{Left-moving fermionic zero-mode}

The transverse wavefunction for the left-moving fermions must satisfy
\begin{eqnarray} (D_{\bar{z}} + \frac{i}{2} Q_{\bar{z}}) f_+ = 0 \qquad \Rightarrow \qquad (\bar{\partial} + \frac{1}{4} \bar{\partial} a + \frac{1}{4} \bar{\partial} \log{\tau_2} ) f_+ = 0 \ . \end{eqnarray}
The general solution is
\begin{eqnarray} f_+(z,\bar{z}) = c(z) \tau_{2}^{-1/4} e^{-a/4} \ , \end{eqnarray}
where $c(z)$ is an arbitrary holomorphic function.  How do we restrict $c(z)$?  For one thing, $\int d^2z \tau_2 e^{a} |f_+|^2$ must be finite, but there is also another condition.  Under a general $SL(2,\mathbbm{Z})$ transformation \eqref{tautrans}, one can verify that
\begin{eqnarray} Q_m \rightarrow Q_m - \partial_m \varphi \ , \quad \textrm{where} \quad \varphi = Arg(c\tau + d) = - \frac{i}{2} \log{ \displaystyle\biggl( \frac{c\tau + d}{c \bar{\tau} + d} \displaystyle\biggr)}. \label{Qtrans} \end{eqnarray}
Since we have gauged $SL(2,\mathbbm{Z})$ in constructing the supergravity background, the equations of motion must transform covariantly.  Thus we require that\footnote{Up to a possible automorphism of the R-symmetry group \cite{KapustinWitten}.  We will come to this point shortly.}
\begin{eqnarray} \psi^i \rightarrow e^{i \varphi/2} \psi^i = \displaystyle\biggl( \frac{c \tau + d}{c \bar{\tau} + d} \displaystyle\biggr)^{1/4} \psi^i \label{fermiontranslaw} \end{eqnarray}
under $SL(2,\mathbbm{Z})$.  This result agrees with the transformation for the $\mathcal{N} = 4$ supersymmetries and fermions derived in \cite{KapustinWitten} by other means.  Also, observe that under $S^2 = -\mathbbm{1}$ we have $\psi \rightarrow i \psi$.  Thus $\psi$ is only invariant under $S^8$--it transforms under a double cover of $SL(2,\mathbbm{Z})$.  It is the same for the dilatino, gravitino and Killing spinor of the supergravity solution.

This can help us in the following way.  In the supergravity background there are some loops that have nontrivial $SL(2,\mathbbm{Z})$ monodromy.  Following the arguments in \cite{Berg} regarding the Killing spinor, when we demand that the equation of motion transform covariantly under $SL(2,\mathbbm{Z})$, what we are really requiring is that the Lorentz monodromy, or holonomy, around these closed loops, due to the nontrivial spin connection, cancel the $SL(2,\mathbbm{Z})$ monodromy.  In other words, the fermions should have trivial holonomy with respect to the total connection $\mathcal{D}_Q$.  We can constrain the zero-mode solution by requiring that it undergo the correct holonomies around various closed loops in the background.

Therefore, what we now need is a detailed picture of the $SL(2,\mathbbm{Z})$ branch cut structure of the background.  The proper way to view the configuration is a generalization of the bottom picture in Figure 3, on page 26 of \cite{Berg}.  Instead of just one $4D7+O7$ system, imagine four of these on the surface of an $S^2$, well separated.  The boxed regions have the charge and monodromy of an $O7^{-}$ plane, and we only consider closed loops that do not intersect them, thus staying in the perturbative framework.  Each set has a point $z_i$ associated to it.  These points are not the locations of any branes, but they have monodromy $S^2$ about them.  In a noncompact case, these points would be taken to infinity.  In the compact case we take them all coincident, so that the net monodromy around this point is $S^8$--ie. trivial.

Thus we are effectively treating each $O7$-region as a point $z_{O7}^{(k)}$, $k = 1,\ldots,4$, which has 5 branches of $SL(2,\mathbbm{Z})$ monodromy emanating from it.  Four branches are lines across which $T$ monodromy occurs and each of these ends on one of the four ``satellite'' $D7$-branes.  Each $D7$-brane is at one of the 24 points $z_{i\infty}^{(n)}$.  One branch is a line across which $S^2$ monodromy occurs, and this line extends to an arbitrary point that we may take far from the $O7$-plane for convenience.  Note that the arrows in the diagram are also important.  The $S^2$ monodromy is associated with a clockwise loop around the $O7$-plane.  A counter-clockwise loop corresponds to $(S^2)^{-1} = S^6$, which is not the same as $S^2$ in the double cover.  Note also that each $O7$-region necessarily swallows up two of the $z_{i\infty}^{(n)}$, so that in this approximation we should take these coincident and equal to the relevant $z_{O7}$.

Therefore, as we encircle any of the $O7$-planes in a clockwise fashion, we require that $\psi \rightarrow i \psi$.  What does this imply for $f_{\pm}$?  Recall that the connection acts on $f_+$ as $(\mathcal{D}_Q)_{\bar{z}} f_+ =0$ and on $f_-$ as $(\mathcal{D}_Q)_z f_- = 0$.  This means that the Lorentz holonomy of $f_+$ should be measured by integrating around the loop with respect to $d\bar{z}$, while the Lorentz holonomy of $f_-$ is measured by integrating around the loop with respect to $dz$.  It follows that we must have $f_+ \rightarrow i f_+$ and $f_- \rightarrow -i f_-$.  This can also be seen by looking at the equation for parallel transport, $t^i D_i \psi = 0$, where $t^i$ is a tangent vector to the curve.

We claim that the unique normalizable solution for $f_+$, with all of the correct $SL(2,\mathbbm{Z})$ monodromies, is obtained by taking
\begin{eqnarray} c(z) & = & \frac{ \eta(\bar{\tau}(z)) }{ \prod_{n=1}^{24} (z - z_{i\infty}^{(n)})^{1/24} } \ .  \end{eqnarray}
Note that $\bar{\tau}$ is a holomorphic function everywhere except at the $z_{i\infty}^{(n)}$ and that $\eta$ is entire on $F_0$.  Thus $\eta(\bar{\tau}(z))$ fails to be holomorphic precisely at the $z_{i\infty}^{(n)}$.  Recall that around these points $\bar{\tau}$ behaves as $\bar{\tau} \sim \frac{i}{2\pi} \log{(z - z_{i\infty}^{(n)})}$, and thus $\eta$ behaves as $(z-z_{i\infty}^{(n)})^{1/24}$.  Therefore $c(z)$ is everywhere holomorphic.  This $c(z)$ yields
\begin{eqnarray} f_+(z,\bar{z}) & = & \frac{1}{\sqrt{\tau_2}} \displaystyle\biggl( \frac{\eta(\bar{\tau} (z)) }{ \bar{\eta}(\tau(\bar{z})) } \cdot \frac{ \prod_{n=1}^{24} (\bar{z} - \bar{z}_{i\infty}^{(n)})^{1/24}}{ \prod_{n=1}^{24} (z - z_{i\infty}^{(n)})^{1/24}} \displaystyle\biggr)^{1/2} \ . \end{eqnarray}
Around clockwise loops enclosing $D7$-branes $f_+$ is invariant, as the phase that $\eta$ acquires under the $T$ transformation is canceled by the phase coming from the explicit $(z - z_{i\infty}^{(n)})$ factors, while $\tau_2$ is invariant under $T$.  For a clockwise rotation around any loop enclosing an $O7$-plane, $f_+$ acquires a phase $f_+ \rightarrow e^{i\pi/2} f_+$.  Where does this phase come from?  The $\eta$ functions and $\tau_2$ are invariant under the $S^2$ monodromy, but when we enclose an $O7$-plane we either enclose 6 of the $z_{i\infty}^{(n)}$ or enclose $6-p$ of them and cross $p$ lines of $T$ monodromy.  In either case, the result of this is a net phase for $f_+$ of $i$.
Finally, it is easy to see that the solution leads to a normalizable mode, as
\begin{eqnarray} \int d^2 z \tau_2 e^{a} |f_+|^2 & = & \int d^2 z e^a = vol_{\mathbbm{C} \mathbbm{P}^1} \ . \end{eqnarray}

We have just argued above that, around an $O7$-plane, the fermions should transform via $f_{\pm} \rightarrow \pm i f_{\pm}$, and we have found a left-moving zero-mode that does precisely this.  But this does not correspond to either the periodic or anti-periodic boundary conditions that we considered in the index calculation.  Fortunately, there is an ambiguity in the $SL(2,\mathbbm{Z})$ transformation of the fermions.  As was pointed out in \cite{KapustinWitten}, the transformation is not unique, but rather can be combined with an automorphism of the supersymmetry algebra.  The Montonen-Olive conjecture states that $SL(2,\mathbbm{Z})$ should commute with the Poincare symmetries, but there is still the global $SU(4)_R$ symmetry.  The automorphism of $SU(4)_R$ must be an inner automorphism, because the classical theory has no symmetry that acts trivially on spacetime and by an outer automorphism of $SU(4)_R$.  Since inner automorphisms of a group are generated by its center, the $SL(2,\mathbbm{Z})$ action is defined up to the action of an element of the center of $SU(4)_R$.  The center of $SU(4)_R$ is generated by the element $\mathcal{J} = diag(e^{i\pi/2},e^{i\pi/2},e^{i\pi/2},e^{i\pi/2})$ which acts as $i$ on the $\mathbf{4}$ of $SU(4)_R$ and as $-i$ on the $\bar{\mathbf{4}}$.

Now, both the left- and right-handed modes are in the $\mathbf{4}$.  Hence, the $SL(2,\mathbbm{Z})$ phases around the $O7$-plane, $\phi = \pm \pi/2$,  are only defined up to $\phi \sim \phi + \pi/2$.  In particular, $ f_+ \rightarrow i f_+$ is \emph{physically equivalent} to\footnote{Note that this \emph{does not} imply that all phases modulo $\pi/2$ are equivalent.  When we compute the $SL(2,\mathbbm{Z})$ monodromy around a closed loop, we get a definite answer (modulo $2\pi$).  This definite answer is physically equivalent to a $\pi/2$ shift of the same answer.  It is \emph{not} physically equivalent to a shift by $\pi$, or $-\pi/2$, etc.  A shift by $\pi$, for example, would be generated by the element $\mathcal{J}^2$ in $SU(4)_R$.  But, as is also discussed in \cite{KapustinWitten}, this corresponds to a $\mathbbm{Z}_2$ fermion number transformation $(-1)^F$.  And this is the $\mathbbm{Z}_2$ element that is already used to extend $SL(2,\mathbbm{Z})$ to its double cover, under which the fermions transform.  Hence this element already commutes with (the double cover of) $SL(2,\mathbbm{Z})$, and can not be used to enlarge the set of physically equivalent phases further.  Similarly a shift by $-\pi/2 \sim 3\pi/2$ would correspond to $\mathcal{J}^3$, but this must be identified with $\mathcal{J}$ and so there is no such shift.}
\begin{eqnarray}   f_+ \rightarrow - f_+ \ .  \end{eqnarray}
Hence, the left-moving mode has anti-periodic boundary conditions.  Therefore it must transform in the anti-symmetric tensor of $O(N_c)$, as claimed.

\subsection{Right-moving fermionic zero-mode}

We will be brief here as the analysis is quite parallel.  The equation of motion for the transverse wavefunction is
\begin{eqnarray} (D_{z} + \frac{i}{2} Q_{z}) f_- = 0 \qquad \Rightarrow \qquad (\partial + \frac{1}{4} \partial a - \frac{1}{4} \partial \log{\tau_2} ) f_- = 0 \ , \end{eqnarray}
with general solution
\begin{eqnarray} f_-(z,\bar{z}) & = & \bar{b}(\bar{z}) \tau_{2}^{1/4} e^{-a/4} \ , \end{eqnarray}
where $b(z)$ is an arbitrary holomorphic function.  The unique normalizable solution with all of the correct $SL(2,\mathbbm{Z})$ monodromies is obtained by taking
\begin{eqnarray} \bar{b}(\bar{z})& = &  \frac{ \bar{\eta}(\tau(\bar{z}))}{ \prod_{n=1}^{24} (\bar{z} - \bar{z}_{i\infty}^{(n)} )^{1/24} } \  , \end{eqnarray}
which yields
\begin{eqnarray} f_-(z,\bar{z}) & = & \displaystyle\biggl( \frac{ \bar{\eta}(\tau(\bar{z})) }{\eta(\bar{\tau} (z)) } \cdot \frac{ \prod_{n=1}^{24} (z - z_{i\infty}^{(n)})^{1/24}}{ \prod_{n=1}^{24} (\bar{z} - \bar{z}_{i\infty}^{(n)})^{1/24}} \displaystyle\biggr)^{1/2} \ . \end{eqnarray}
As we move around a clockwise loop enclosing an O7-plane, we have $f_- \rightarrow -i f_-$.  This solution leads to a normalizable mode,
\begin{eqnarray} \int \tau_2 e^a |f_-|^2 = \int \tau_2 e^a < \infty \ , \end{eqnarray}
since $\tau_2(z,\bar{z})$ is a function with only logarithmic singularities on $\mathbbm{C} \mathbbm{P}^1$.  Finally, the monodromy $f_- \rightarrow -i f_-$ is physically equivalent to
\begin{eqnarray} f_- \rightarrow f_- \ . \end{eqnarray}
Hence, the right-moving mode has periodic boundary conditions and therefore must transform in the symmetric tensor of $O(N_c)$.

\subsection{The 3-3 scalars}

The quadratic action for the $D3$-brane scalars $M^{ij}$ is
\begin{eqnarray} S_{\textrm{3-3},(quad)}[M^{ij}] & = & - \frac{1}{4\pi} \int d^4 x \sqrt{-g} \textrm{tr} \displaystyle\biggl( \partial_m M_{ij} \partial^m M^{ij} \displaystyle\biggr), \end{eqnarray}
which leads to the equations of motion
\begin{eqnarray} \partial_m ( \sqrt{-g} g^{mn} \partial_n M^{ij}) =  0 \qquad \Rightarrow \qquad (e^{a} \partial_\mu \partial^\mu + 4 \partial \bar{\partial}) M^{ij}  =  0 \ . \end{eqnarray}
Clearly massless modes correspond to $\partial \bar{\partial} M^{ij} = 0$.  The general solution is a sum of holomorphic and anti-holomorphic functions.  Since we are on a compact space, and the equation of motion has no $\tau$ dependence, there exists precisely one normalizable and modular invariant solution: the constant.  This is periodic as we go around the $O7$-plane, and therefore it must be in the symmetric tensor of $O(N_c)$.  This gives us six real scalars in the symmetric tensor representation.  We expect two more scalars in order to complete the $\mathbf{8}_v$ of $SO(8)_R$ in the Type I dual theory, that describes the transverse fluctuations of the $D1$-string.  We also expect zero-modes that represent a $1+1$-dimensional gauge field in the anti-symmetric tensor representation.  All of these remaining modes must come from the 3-3 gauge field.

\subsection{The 3-3 gauge field}

The quadratic action for the $D3$-brane gauge field is
\begin{eqnarray} S_{3-3,(quad)}[A_m] & = & - \frac{1}{8\pi} \int d^4 x \sqrt{-g} \textrm{tr} \displaystyle\biggl( \tau_2 F_{mn} F^{mn} + \frac{1}{2} \tau_1 \epsilon^{mnpq} F_{mn} F_{pq} \displaystyle\biggr), \end{eqnarray}
where $\epsilon^{0123} = (-g)^{-1/2}$, and we only need to consider the $U(1)$ part of the field strength: $F_{mn} = \partial_m A_n - \partial_n A_m + O(A^2)$.  We integrate by parts in order to put the Lagrangian in the form $A_m \mathcal{O}^{mn} A_n$.  Doing so, one finds the usual curved space Maxwell operator and, additionally, terms that are proportional to spacetime derivatives of $\tau_1,\tau_2$.  After choosing the usual covariant Lorentz gauge, $D^m A_m = 0$, the action can be put in the form
\begin{eqnarray} S & = & \frac{1}{4\pi} \int d^4 x \tau_2 \sqrt{-g} \textrm{tr} A_m \mathcal{O}^{mn} A_n \ , \qquad \textrm{with} \label{GaugeAction} \\
\mathcal{O}^{mn} & = &  g^{mn} D_p D^p - R^{mn} + g^{mn} \frac{\partial_p \tau_2}{\tau_2} D^p + \frac{ \partial^m \tau_2}{\tau_2} D^n - \frac{ D^n \tau_2}{\tau_2} D^m + \nonumber \\
& & - \frac{D_p \tau_1}{\tau_2} \epsilon^{mnpq} \partial_q \ . \end{eqnarray}
(The term $\partial^m \log{\tau_2} D^n$ gives zero when acting on $A_n$, but we choose to keep it because it will simplify the form of the operator).

We work in lightcone coordinates $x^{\pm} = x^0 \pm x^1$, so that the metric is given by $ds^2 = - dx^+ dx^- + e^{a(z,\bar{z})} dz d\bar{z}$.  After some work one eventually finds the following expression:
\begin{eqnarray} \mathcal{O}^{mn} & = & \mathcal{O}_{1+1}^{mn} + \mathcal{O}_{t}^{mn}(z,\bar{z}), \end{eqnarray}
where
\begin{eqnarray} (\mathcal{O}_{1+1}^{mn}) & = & 8 \left(\begin{array}{c c c c} 0 & \partial_+ \partial_- & \bar{\partial} \log{\tau_2} \partial_- & \partial \log{\tau_2} \partial_- \\ \partial_+ \partial_- & 0 & 0 & 0 \\ -\bar{\partial} \log{\tau_2} \partial_- & 0 & 0 & -e^{-a} \partial_+ \partial_- \\ -\partial \log{\tau_2} \partial_- & 0 & - e^{-a} \partial_+ \partial_- & 0 \end{array}\right) \end{eqnarray}
and
\begin{eqnarray} (\mathcal{O}_{t}^{mn}) & = &
 8 e^{-a} \left(\begin{array}{c c c c} 0 & \begin{array}{c} - (\partial \bar{\partial} + \partial \log{\tau_2} \bar{\partial} \\ + \bar{\partial} \log{\tau_2} \partial) \end{array}  & 0 & 0 \\ - \partial \bar{\partial} & 0 & 0 & 0 \\ 0 & 0 & 0 & \begin{array}{c} e^{-a} (\bar{\partial} \partial - \bar{\partial} a \partial  \\ + \bar{\partial} \log{\tau_2} \partial ) \end{array} \\ 0 & 0 &  \begin{array}{c} e^{-a} (\partial \bar{\partial} - \partial a \bar{\partial} \\ + \partial \log{\tau_2} \bar{\partial} ) \end{array} & 0 \end{array}\right).  \nonumber \\ \end{eqnarray}
In deriving these expressions we have used the fact that $\tau(\bar{z})$ is an anti-holomorphic function so, for instance, $\partial \tau_1 = -i \partial \tau_2$ etc.  Note that the order of the columns and rows in these matrices is $(+,-,z,\bar{z})$.

Zero modes are obtained as solutions of
\begin{eqnarray} \mathcal{O}_{t}^{mn} A_n = 0 \ . \label{ZeroModeEqns} \end{eqnarray}
Note we must also make sure that the gauge constraint
\begin{eqnarray} D^n A_n =  - 2 ( \partial_+ A_- + \partial_- A_+ ) + 2 e^{-a} (\partial A_{\bar{z}} + \bar{\partial} A_z ) = 0 \end{eqnarray}
can be maintained.  We make the ansatz
\begin{eqnarray} A_m(x^\mu, z ,\bar{z}) = a_m (x^\mu) \Psi_m (z,\bar{z}) \qquad \textrm{(no sum)}. \end{eqnarray}
Then the equations \eqref{ZeroModeEqns} are
\begin{eqnarray} \partial \bar{\partial} \Psi_+ & = & 0 \ ,    \label{PsiplusEq} \\
(\partial \bar{\partial} + \partial \log{\tau_2} \bar{\partial} + \bar{\partial} \log{\tau_2} \partial ) \Psi_- & = & 0 \ ,  \label{PsiminusEq} \\
(\partial - \partial a + \partial \log{\tau_2} ) \bar{\partial} \Psi_z & = & 0 \ , \label{PsizEq} \\
(\bar{\partial} - \bar{\partial} a + \bar{\partial} \log{\tau_2} ) \partial \Psi_{\bar{z}} & = & 0 \ . \label{PsibarzEq} \end{eqnarray}
Assuming that these equations are satisfied, it is easy to show that the gauge field action \eqref{GaugeAction} reduces to
\begin{eqnarray} S & = & \frac{2}{\pi} \int d^2 z \tau_2 \sqrt{-g} \Psi_+ \Psi_- \int d^2 x \displaystyle\biggl( a_+ \partial_+ \partial_- a_- + a_- \partial_+ \partial_- a_+ \displaystyle\biggr) + \nonumber \\
& & - \frac{2}{\pi} \int d^2 z \tau_2 \sqrt{-g} e^{-a} \Psi_z \Psi_{\bar{z}} \int d^2 x \displaystyle\biggl( a_z \partial_+ \partial_- a_{\bar{z}} + a_{\bar{z}} \partial_+ \partial_- a_z \displaystyle\biggr) + \nonumber \\
& & + \frac{2}{\pi} \int d^2 z \tau_2 \sqrt{-g} \partial \log{\tau_2} \Psi_+ \Psi_{\bar{z}} \int d^2 x \displaystyle\biggl( a_+ \partial_- a_{\bar{z}} - a_{\bar{z}} \partial_- a_+ \displaystyle\biggr) + c.c.  \end{eqnarray}

In order for this to give an effective $1+1$-dimensional action for massless modes, we require that the integrals over the transverse space be finite and, additionally, the third term must vanish.  In order for the third term to vanish, there are two disjoint possibilities:
\begin{eqnarray} \Psi_z = \Psi_{\bar{z}} = 0 \qquad \textrm{or} \qquad \Psi_+ = 0 \ . \end{eqnarray}
These will lead to two different sets of zero-modes.  We investigate each separately.

\subsubsection{The $A_{\pm}$ zero-modes}

This solution corresponds to setting
\begin{eqnarray} \Psi_{z} = \Psi_{\bar{z}} = 0 \ . \end{eqnarray}
The reality of the gauge field $A_m$ implies that $\phi_{\pm}, \Psi_{\pm}$ may be taken real.  The most general real solutions to the $\Psi_{\pm}$ equations \eqref{PsiplusEq}, \eqref{PsiminusEq} are
\begin{eqnarray} \Psi_+(z,\bar{z}) = \psi_+(z) + \bar{\psi}_+(\bar{z}), \qquad \Psi_-(z,\bar{z}) = \frac{1}{\tau_2} (\psi_-(z) + \bar{\psi}_-(\bar{z}) ), \end{eqnarray}
where $\psi_{\pm}(z)$ are arbitrary holomorphic functions of $z$ (possibly involving $\bar{\tau}(z)$).  Again, there are two requirements that will select out a unique solution: the integral over the transverse space must be finite, and around closed loops $\Psi_{\pm}$ should undergo monodromies consistent with their transformations under the corresponding $SL(2,\mathbbm{Z})$ actions.

This brings us to the question of how do $\Psi_{\pm}$ transform under a general $\Lambda \in SL(2,\mathbbm{Z})$.  As in the case of the fermions, we answer this by requiring that the equations of motion transform covariantly under $SL(2,\mathbbm{Z})$.  First consider the $\Psi_+$ equation of motion \eqref{PsiplusEq}.  The operator is invariant under $SL(2,\mathbbm{Z})$ and so we require $\Psi_+$ to be invariant.  Thus $\psi_+$ is a holomorphic function of $z$ only and can not depend on $\bar{\tau}$.  Now consider
\begin{eqnarray} (\partial \bar{\partial} + \partial \log{\tau_2} \bar{\partial} + \bar{\partial} \log{\tau_2} \partial ) \Psi_- =  0 \ . \end{eqnarray}
Under general $\Lambda = $ {\scriptsize$\begin{pmatrix} a&b \cr c&d \end{pmatrix}$}, the differential operator transforms and it is not at all obvious how $\Psi_-$ can transform in order to counter these new terms, and not introduce any others.

We need to write the $\Psi_-$ equation in an $SL(2,\mathbbm{Z})$ covariant fashion, involving the $U(1)$ Kahler connection.  It is straightforward to show that \eqref{PsiminusEq} can be written as
\begin{eqnarray} (D^i - q^\ast \epsilon^{ij} Q_j )(D_i + q \epsilon_{ik} Q^k ) \Psi_- = 0 \ ,  \qquad \textrm{with} \qquad q = 2i \ , \label{CovPsiminusEq} \end{eqnarray}
where $i,j = z,\bar{z}$ run over the transverse coordinates and $\epsilon_{ij}$ is the Levi-Cevita tensor density on the transverse 2-manifold.  Our conventions will be $\epsilon_{z\bar{z}} = - \epsilon_{\bar{z} z} = - \sqrt{g}$, from which it follows that $\epsilon^{z \bar{z}} = - \epsilon^{\bar{z} z} = g^{-1/2}$.  (If this sign is flipped, then the charge $q = 2i$ must also change sign).  In this form we can now determine the proper transformation of $\Psi_-$ under $SL(2,\mathbbm{Z})$.  Recalling the transformation law of $Q_i$, \eqref{Qtrans}, one has
\begin{eqnarray} \Delta 2 i \epsilon_{ij} Q^j & = & -2 i \epsilon_{ij} \partial^j \varphi = \left\{ \begin{array}{c c} 2 i \partial_z \varphi, & i = z \\ - 2i \partial_{\bar{z}} \varphi, & i = \bar{z} \end{array}\right.  \nonumber \\
& = & \left\{ \begin{array}{c c} - \partial \log{(c\bar{\tau} + d)}, & i = z \\ - \bar{\partial} \log{(c\tau + d)}, & i = \bar{z} \end{array}\right. .\end{eqnarray}
Hence we require that
\begin{eqnarray} \partial_i \Psi_- \rightarrow \partial_i \Psi_- + \left\{ \begin{array}{c c}  \partial \log{(c\bar{\tau} + d)}, & i = z \\  \bar{\partial} \log{(c\tau + d)}, & i = \bar{z} \end{array}\right. .\end{eqnarray}
This is achieved with the transformation
\begin{eqnarray} \Psi_- \rightarrow e^{\log{(c\tau+d)} + \log{(c\bar{\tau}+d)}} \Psi_- = (c\tau + d) (c \bar{\tau} + d) \Psi_- \ . \label{Psiminustrans} \end{eqnarray}
In other words, $\Psi_-$ should transform as a modular form of weight $(1,1)$.  As a check, one could now go back to the original equation \eqref{PsiminusEq} and see that it does indeed transform covariantly using this transformation rule.  After considerable algebra and some nontrivial cancellations, one finds that it actually works.

This transformation is consistent with $\Psi_- = 1/\tau_2$ and therefore we require that $\psi_-(z)$ be independent of $\bar{\tau}(z)$.  Now we use the requirement of normalizability to fix the functions $\psi_{\pm}(z)$.  The guage field action  evaluated on these zero-modes is
\begin{eqnarray} S & = & \frac{2}{\pi} \int i d^2 z e^{a} Re(\psi_+(z)) Re(\psi_-(z)) \int d^2 x \displaystyle\biggl( a_+ \partial_+ \partial_- a_- + a_- \partial_+ \partial_- a_+ \displaystyle\biggr). \end{eqnarray}
Since $e^a \sim 1/|z|^4$ as $z \rightarrow \infty$, the only functions $\psi_{\pm}$ that are \emph{entire} on $\mathbbm{C}$ and give a \emph{non-zero}, \emph{finite} result for the integral are
\begin{eqnarray} \psi_{\pm} = const \ . \end{eqnarray}
Thus we arrive at precisely two real massless zero-modes $a_{\pm}$, that transform as a vector under the $SO(1,1)$ Lorentz group.

Observe that the transverse wavefunctions
\begin{eqnarray} \Psi_+(z,\bar{z}) = const \ , \qquad \Psi_-(z,\bar{z}) = \frac{const}{\tau_2} \end{eqnarray}
are periodic around the $O7$-planes.  Indeed, the $SL(2,\mathbbm{Z})$ transformations for these wavefunctions indicate that they should be invariant under $S^2$.  However, recall that $S^2$ acts on the string worldsheet as $\Omega$, worldsheet orientation reversal.  In addition to flipping the Chan-Paton indices, this gives a minus sign when acting on the string modes corresponding to the gauge field.  Thus, in order for the overall state to be single-valued around the $O7$-planes, these zero-modes must transform in the \emph{anti-symmetric tensor} of $O(N_c)$.  Thus, as one would expect, the zero-modes of the $(A_+,A_-)$ components of the 3-3 gauge field correspond to the $1+1$-dimensional gauge field on the intersection.

Finally, although the above $\Psi_{\pm}$ satisfy the equations of motion, we must still check that the gauge condition can be maintained.  With $\Psi_z,\Psi_{\bar{z}}$ set equal to zero, we have
\begin{eqnarray} D^m A_m & = & -2 (\partial_+ a_- \Psi_- + \partial_- a_+ \Psi_+) = 0 \ , \end{eqnarray}
which is just the standard Lorentz gauge condition in $1+1$ dimensions.  Since $\Psi_+,\Psi_-$ are linearly independent functions of $z$, the gauge condition can only be satisfied by taking
\begin{eqnarray} \partial_+ a_- = \partial_- a_+ = 0 \label{reducedgaugecondition} \ . \end{eqnarray}
If we were dealing with a $U(1)$ gauge field this would imply that the field strength vanishes.  However, in the non-Abelian case there is the term $F_{+-}^{(2)} \sim a_+ a_-$.  It follows from \eqref{reducedgaugecondition} that this corresponds to a non-propagating electric field along the intersection, which is, of course, what one would expect.

\subsubsection{The $A_z,A_{\bar{z}}$ zero-modes}

Now let us suppose that
\begin{eqnarray} \Psi_+ = 0 \ . \end{eqnarray}
The equations \eqref{PsizEq}, \eqref{PsibarzEq} are consistant with taking $(\Psi_z)^\ast = \Psi_{\bar{z}}$, as they should be since the reality of the gauge field requires it.  Now, recall from the supergravity solution that
\begin{eqnarray} e^{a(z,\bar{z})} = \tau_2 g(z) \bar{g}(\bar{z}) \ , \qquad \textrm{where} \qquad g(z) \equiv \frac{ \eta^2(\bar{\tau}(z)) }{\prod_{n=1}^{24} (z - z_{i\infty}^{(n)})^{1/12} } \ . \end{eqnarray}
Hence, $\partial (a - \log{\tau_2}) = \partial \log{g}$ is a holomorphic function and the operators $(\partial - \partial a + \partial \log{\tau_2})$, $\bar{\partial}$ commute.  Thus the general solution for $\Psi_z,\Psi_{\bar{z}}$ is given by
\begin{eqnarray} \Psi_z(z,\bar{z}) = \bar{\alpha}(\bar{z}) g(z) + \beta(z) \ , \qquad \Psi_{\bar{z}} = \alpha(z) \bar{g}(\bar{z}) + \bar{\beta}(\bar{z}) \ , \end{eqnarray}
where $\alpha(z),\beta(z)$ are arbitrary holomorphic functions.  We can quickly eliminate one of these functions by considering the question of normalizability.  If $\beta(z)$ is nonzero, then the integral over the transverse space will contain a term $\int \tau_2 |\beta |^2$.  Since $\tau_2 \rightarrow const >0$ as $z \rightarrow \infty$, it is clear that this can not be normalizable for any function $\beta$ entire on $\mathbbm{C}$.  Therefore we may take
\begin{eqnarray} \beta = 0 \ . \end{eqnarray}
In order to fix $\alpha(z)$, however, we will need to consider how $\Psi_z,\Psi_{\bar{z}}$ transform under $SL(2,\mathbbm{Z})$.

One can check that the following Lorentz and $SL(2,\mathbbm{Z})$ covariant equation of motion reduces to \eqref{PsizEq}:
\begin{eqnarray} ( D^i - q^\ast ( \epsilon^{ij} - g^{ij} ) Q_j )( D_i + q(\epsilon_{ik} - g_{ik}) Q^k ) \Psi_z = 0 \ , \qquad \textrm{with} \quad q = i \ . \label{CovPsizEq} \end{eqnarray}
Using this we can derive the transformation law of $\Psi_z$ under $SL(2,\mathbbm{Z})$.  Under $\tau \rightarrow \Lambda \tau$ we have
\begin{eqnarray} \Delta i (\epsilon_{ij} - g_{ij}) Q^j & = & - i (\epsilon_{ij} - g_{ij} ) \partial^i \varphi = \left\{ \begin{array}{c c} 2 i \partial_z \varphi, & i = z \\ 0, & i = \bar{z} \end{array}\right. \nonumber \\
& = & \left\{ \begin{array}{c c} - \partial \log{(c \bar{\tau} + d)}, & i = z \\ 0, & i = \bar{z} \end{array}\right. . \end{eqnarray}
Therefore we require
\begin{eqnarray} \Psi_z \rightarrow e^{\log{(c\bar{\tau} + d)}} \Psi_z = (c \bar{\tau} + d) \Psi_z \ . \label{Psiztrans} \end{eqnarray}
For $\Psi_{\bar{z}}$ the manifestly covariant equation of motion is
\begin{eqnarray}  ( D^i + i( \epsilon^{ij} + g^{ij} ) Q_j )( D_i + i(\epsilon_{ik} + g_{ik}) Q^k ) \Psi_z  & = & 0 \ , \end{eqnarray}
which leads to the expected transformation rule
\begin{eqnarray} \Psi_{\bar{z}} \rightarrow (c \tau + d) \Psi_{\bar{z}} \ . \end{eqnarray}

These rules mean that $\Psi_z,\Psi_{\bar{z}}$ should be invariant under $T$ and change sign under $S^2$.  Therefore, they should be antiperiodic around the $O7$-planes.  The function $g(z)$ (and its conjugate) has just these properties, reasoning along the same lines as we did for the fermionic zero-mode solutions.  Thus we may take $\alpha(z) = const$ so that
\begin{eqnarray} \Psi_z(z,\bar{z}) = \frac{\eta^2(\bar{\tau}(z)) }{ \prod_{n=1}^{24} (z - z_{i\infty}^{(n)} )^{1/12} } \ , \qquad \Psi_{\bar{z}}(z,\bar{z}) = \frac{\bar{\eta}^2(\tau(\bar{z})) }{ \prod_{n=1}^{24} (\bar{z} - \bar{z}_{i\infty}^{(n)} )^{1/12} } \ . \end{eqnarray}
To see that $\alpha = const$ is the only solution, observe that
\begin{eqnarray} \int \tau_2 \Psi_z \Psi_{\bar{z}} = \int \tau_2 |g|^2  |\alpha|^2 = \int e^a |\alpha|^2 \ . \end{eqnarray}
Then by the same arguments as in the last section, we must have $\alpha = const$ for normalizability.

When we combine the anti-periodicity of the transverse wavefunction with the extra minus sign coming from the action of $\Omega$ on these string modes, we see that the $A_z,A_{\bar{z}}$ zero-modes must transform in the \emph{symmetric} tensor representation of $O(N_c)$.  Hence they have the right quantum numbers to be the remaining two scalars that complete the $SO(8)_v$ transverse fluctuation modes of the $1+1$-dimensional theory.  This is what we would expect; $T$-duality in a given direction maps the component of the gauge field in that direction to a scalar representing transverse fluctuations.

Note that for this solution we set $\Psi_+ = 0$, but it was not necessary to set $\Psi_-$ to zero.  Since $\partial_z \Psi_{\bar{z}} = \partial_{\bar{z}} \Psi_z = 0$, the gauge condition reduces to $\partial_+ A_- = 0$.  Thus we may either have $\Psi_- = 0$ or $\partial_+ a_- = 0$.  This is pure gauge from the $1+1$-dimensional point of view.

\end{document}